\documentclass[reqno,12pt]{amsart}

\usepackage{amsthm,amsmath,amssymb,amsfonts}
\usepackage{fullpage}

\usepackage{graphicx}
\usepackage{float}
\usepackage{subfig}

\begin{document}

\newtheorem{thm}{Theorem}[section]
\newtheorem{cor}{Corollary}[section]
\newtheorem{lem}{Lemma}[section]
\newtheorem{prop}{Proposition}[section]
\newtheorem{rem}{Remark}[section]

\def\Ref#1{Ref.~\cite{#1}}

\def\const{\text{const.}}
\def\Rnum{{\mathbb R}}
\def\sgn{{\rm sgn}}
\def\sech{{\rm sech}}
\def\cn{{\rm cn}}
\def\arccn{{\rm arccn}}
\def\sn{{\rm sn}}
\def\arcsn{{\rm arcsn}}
\def\K{{\rm K}}

\def\Dop{{\mathcal{D}}}
\def\X{{\rm X}}
\def\Y{{\rm Y}}

\def\k{k}
\def\q{q}
\def\s{\sigma}
\def\V{\mathcal{V}}

\def\KP{{\it KP}}
\def\c{{\rm c}}

\tolerance=20000
\allowdisplaybreaks[3]

\title{Nonlinearly dispersive KP equations\\ with new compacton solutions}

\author{
S.C. Anco${}^1$
\lowercase{\scshape{and}}
M.L. Gandarias${}^2$
\\\\
${}^1$D\lowercase{\scshape{epartment}} \lowercase{\scshape{of}} M\lowercase{\scshape{athematics and}} S\lowercase{\scshape{tatistics}}\\
B\lowercase{\scshape{rock}} U\lowercase{\scshape{niversity}}\\
S\lowercase{\scshape{t.}} C\lowercase{\scshape{atharines}}, ON L2S3A1, C\lowercase{\scshape{anada}} \\\\
${}^2$D\lowercase{\scshape{epartment}} \lowercase{\scshape{of}} M\lowercase{\scshape{athematics}}\\
F\lowercase{\scshape{aculty of}} S\lowercase{\scshape{ciences}}, U\lowercase{\scshape{niversity of}} C\lowercase{\scshape{\'adiz}}\\
P\lowercase{\scshape{uerto}} R\lowercase{\scshape{eal}}, C\lowercase{\scshape{\'adiz}}, S\lowercase{\scshape{pain}}, 11510\\
}

\date{}

\begin{abstract}
A complete classification of compacton solutions is carried out for
a generalization of the Kadomtsev-Petviashvili (KP) equation
involving nonlinear dispersion in two and higher spatial dimensions.
In particular, precise conditions are given on the nonlinearity powers in this equation
under which a travelling wave can be cut off to obtain a compacton.
Numerous explicit examples having various wave profiles are derived, 
including a quadratic function,
powers of a cosine,
and powers of Jacobi $\cn$ functions,
all of which are symmetric.
The cosine and $\cn$ symmetric compactons have an anti-symmetric counterpart. 
In comparison, explicit solitary waves of the generalized KP equation are found to have
profiles given by a power of a sech and a reciprocal quadratic function.
Kinematic properties of all of the different types of compactons and solitary waves
are discussed,
along with conservation laws of the generalized KP equation.
\end{abstract}


\maketitle

\section{Introduction}\label{sec:intro}

The phenomenon of spatial localization of energy
occurs in numerous physical phenomena involving nonlinearity and dispersion.
Prominent examples are solitons,
which have attracted a lot of attention since their discovery 50 years ago.
A soliton is a travelling wave whose profile has an exponential localization that is preserved in nonlinear interactions.
The exponential tail of solitons indicates that the underlying nonlinear dispersive wave equation exhibits an infinite propagation speed for disturbances.

Wave equations in which the dispersive terms themselves are nonlinear
allow disturbances to propagate with a finite speed.
Such equations arise in a variety of physical models and possess travelling waves whose profile has compact spatial support \cite{RosHym,Ros,Ros2000,RosOro}. 
These truly localized waves are known as compact waves or compactons,
and for some equations they preserve their compactness in nonlinear interactions \cite{RosHym,Ros,ZilRos}.
A compacton is a classical (strong) solution but, in contrast to a soliton, 
it is typically not smooth. 

The simplest soliton equation, 
combining weak nonlinearity and dispersion in one spatial dimension, 
is the well-known Korteweg-de Vries (KdV) equation.
Compactons were first found for a nonlinearly-dispersive generalization of the KdV equation
known as the $K(m,n)$ equation \cite{RosHym},
$u_t+a(u^m)_x+b(u^n)_{xxx}=0$,
where $m,n>0$ are the nonlinearity powers,
and $a,b\neq0$ are the coefficients.
This equation possesses cosine-type compacton solutions when $1<m=n\leq3$.
The case $K(2,2)$ arises from a continuum limit of an anharmonic lattice chain 
with nonlinear dispersion \cite{RosHym}. 

In two and three spatial dimensions,
two versions of the KdV equation arise in physical models
where waves have a small transverse component in their motion.
One is the Zakharov-Kuznetsov (ZK) equation \cite{ZakKuz},
and the other is the Kadomtsev-Petviashvili (KP) equation \cite{KadPet,KuzMus,KarBel}.

A nonlinearly-dispersive generalization of the ZK equation, 
known as the $C_N(m,p,n)$ equation, is given by \cite{RosHymSta}
$u_t+a (u^m)_x+ b(u^p(\Delta u^n)_x)=0$
in terms of the three-dimensional Laplacian $\Delta$,
where the ZK equation is the special case with nonlinear powers $m=2$, $n=1$, $p=0$.
In \Ref{Ros2000,Ros2006},
explicit compacton solutions have been derived,
and an extensive numerical study of compacton interactions has been reported
in \Ref{RosHymSta}.
See \Ref{RosZil2018} for a survey of results.

An interesting question is to consider the analogous generalization of the $K(m,n)$ equation:
\begin{equation}\label{Ndim.KPmn}
(u_t+a(u^m)_x +b(u^n)_{xxx})_x +\s\Delta_\perp u =0
\end{equation}
with nonlinearity powers $m,n>0$, 
and coefficients $a,b\neq0$, $\s^2=1$,
where $\Delta_\perp$ is the Laplacian transverse to $x$
in $N>1$ spatial dimensions.
We will call this the \emph{$K_N(m,n)$ equation}. 
It reduces to the $K(m,n)$ equation when $N=1$
and to the KP equation when $m=2$, $n=1$, $N=2$.
A special case $m=n=2$, $N=2$ was briefly considered in \Ref{RosHym}.
The general case when $N=2$ 
will be called the \emph{$\KP(m,n)$ equation} for simplicity.
Similarly to the $K(2,2)$ equation, 
the case $\KP(2,2)$ should describe the continuum limit of an anharmonic lattice 
with nonlinear dispersion and weak transversality. 

The KP equation in two dimensions possesses line solitary waves, 
and in three dimensions, plane solitary waves.
Correspondingly,
a \emph{line compacton} is a travelling wave of the form 
\begin{equation}\label{travellingwave}
u=U(\xi),
\quad
\xi = x +\mu y - \nu t
\end{equation}
such that the function $U$ has compact support.
The parameters $(\mu,\nu)$ determine both the direction and the speed of the wave.
Specifically, $c=\nu/\sqrt{1+\mu^2}$ is the speed,
and $\theta=\arctan(\mu)$ is the direction angle with respect to the positive $x$ axis.
A \emph{plane compacton} in $N=3$ dimensions has a similar form
in terms of a compact function
\begin{equation}\label{3dim.travellingwave}
u=U(\xi),
\quad
\xi = x +\mu_1 y + \mu_2 z - \nu t
\end{equation}
with a speed given by $c=\nu/\sqrt{1+\mu_1{}^2+\mu_2{}^2}$, 
a polar angle given by $\theta=\arctan(\mu_1)$ in the $x$-$y$ plane,
and an azimuthal angle given by $\phi=\arctan(\mu_2)$ with respect to the $z$ axis.
Note that line waves in two dimensions have
a symmetry characterization of being invariant under two translation symmetries
$\Y = \nu \partial_x +\partial_t$ and $\X = \mu \partial_x -\partial_y$,
while in $N=3$ dimensions a similar characterization for plane waves is given by
the three translation symmetries
$\Y$, $\X_1 = \mu_1 \partial_x -\partial_y$, $\X_2 = \mu_2 \partial_x -\partial_z$.

Travelling wave solutions of the $K_N(m,n)$ equation in any dimension $N\geq1$
are determined by a fourth-order ODE for $U$.
The ODE is straightforward to integrate twice,
which yields a reduction to a second-order ODE for $U$.
This reduced ODE can be expressed as a forced nonlinear oscillator for $V=U^n$
with two integration constants.
There are two cases in which further integration is possible. 
Firstly, when one of the constants is zero such that the ODE is translation invariant, 
a quadrature can be obtained for $U$. 
Secondly, when $m=n$, the reduced ODE in the case $\nu =\s |\mu|^2$ becomes linear,
which leads to a complete integration giving the explicit solution for $U$.
The integrations can be carried out in a simple way 
by use of a symmetry multi-reduction method \cite{AncGan2020}
applied to the set of translation symmetries that characterize travelling waves
in $N$ dimensions.
The method directly yields all conservation laws of the $K_N(m,n)$ equation that
are invariant under these symmetries,
including any special cases for $m,n$ in which additional invariant conservation laws are admitted.
Invariance turns out to imply that each conservation law reduces for $u=U(\xi)$
to a first integral of the travelling wave ODE,
from which the ODE is reduced to quadrature through combining the first integrals.

The main goals in this paper are 
to find systematically the compacton solutions of the $K_N(m,n)$ equation 
and to examine their kinematic features. 
Convex nonlinearities, $\sgn(a)=\sgn(b)$,
and non-convex nonlinearities, $\sgn(a)=-\sgn(b)$, will be considered;
the corresponding powers $m$ and $n$ will be positive,
including the linear dispersive case $n=1$ 
but excluding linear convectivity $m=1$. 
Compactons will be shown to be supported for sublinear powers $m<1$ 
as well as superlinear powers $m>1$. 
The precise sense in which a compacton is a classical (strong) solution will be discussed. 
In particular, this involves working with the equation (and travelling wave ODE)
in a non-singular form.
In this setting, 
compactons arise from travelling waves $U(\xi)$ on a domain $L\leq\xi\leq L$ 
with the property that $U$ and its derivatives $U',\ldots,U''''$ vanish at the endpoints,
allowing $U$ to be extended to be $0$ for all $|\xi|\geq L$. 

Several types of compactons are obtained in an explicit form.
Their profiles $U(\xi)$ are given by powers of a variety of different expressions: \\
$\bullet$
cosine and sine of $\xi$ \\
$\bullet$ 
Jacobi $\cn$ and $\sn$ of $\xi$ \\
$\bullet$
quadratic polynomial in $\xi$ \\
$\bullet$ 
linear in both sine and cosine of $\xi$ with coefficients involving $\xi$\\
The latter type is new compared to the types of compactons that are known for the $K(m,n)$ equation and has the novel feature of being non-symmetric.
The former types come in symmetric and antisymmetric sign-changing versions,
some of which are also new. 

The rest of the paper is organized as follows. 
In section~\ref{sec:travellingwaves},
the reduction of the travelling wave ODE in $N=2$ dimensions is explained
and all solitary wave solutions are obtained.
In section~\ref{sec:conditions},
the conditions under which a travelling wave in $N=2$ dimensions
can be cut off to obtain a compacton as a classical (strong) solution are summarized. 
All compactons in the case $m=n$, $\nu=\s|\mu|^2$ are obtained in section~\ref{sec:linearcase},
and their kinematical features are discussed.
In section~\ref{sec:nonlinearcase}, 
explicit compactons in the cases $m\neq n$ and $\nu\neq \s|\mu|^2$
are presented having the forms just noted. 
Their kinematics features are also discussed.
These results are extended to $N\geq3 $ dimensions in section~\ref{sec:Ndim}.

Some concluding remarks, 
as well as applications to physical phenomena in $N=2,3$ dimensions, 
are made in section~\ref{sec:conclude}.

\section{Travelling wave ODE}\label{sec:travellingwaves}

In $N=2$ dimensions, the $K_N(m,n)$ equation \eqref{Ndim.KPmn} has the form
\begin{equation}\label{KPmn}
(u_t+a(u^m)_x+b(u^n)_{xxx})_x+\s u_{yy} =0 .
\end{equation}
Travelling wave solutions \eqref{travellingwave} are given by
the fourth-order ODE
\begin{equation}\label{U.ODE}
( b (U^n)'' +a U^m -\k U )'' =0 
\end{equation}
with
\begin{equation}\label{lin.coeff}
\k =\nu -\s\mu^2 . 
\end{equation}
Two integrations directly give the second-order ODE
\begin{equation}\label{U.reducedODE}
b (U^n)'' +a U^m -\k U = C_1 \xi +C_2
\end{equation}
where $C_1,C_2$ are arbitrary constants.
It can be simplified by the change of variable
\begin{equation}\label{V.rel}
U = V^{1/n} ,
\end{equation}
which leads to the equivalent ODE
\begin{equation}\label{V.reducedODE}
b V'' + a V^{m/n} -\k V^{1/n} = C_1 \xi +C_2
\end{equation}
for $V(\xi)$.
This has the form of a forced nonlinear oscillator equation.
There are two cases in which it can be solved explicitly up to quadrature,
as mentioned in Section~\ref{sec:intro}. 

In the case $C_1=0$,
the ODE \eqref{V.reducedODE} is invariant under translation in $\xi$,
and hence its solution is given by the quadrature
\begin{equation}\label{V.quadrature}
\int \frac{dV}{\sqrt{ \tfrac{C_3}{2b} + \tfrac{C_2}{2b} V +\tfrac{\k n}{2(n+1)b} V^{1+1/n} - \tfrac{na}{2(m+n)b} V^{1+m/n} }}
= \pm \xi + C_4
\end{equation}
where $C_3,C_4$ are arbitrary constants.
This yields profiles $V(\xi)$ that are symmetric about their peak. 

In the case $C_1\neq 0$,
observe when $m=n$ and $\k=0$ that the ODE \eqref{V.reducedODE}
becomes linear
\begin{equation}\label{V.reducedODE.lin}
a V +b V'' = C_1 \xi +C_2 .
\end{equation}
Its solution is given by
\begin{equation}\label{V.lin}
V(\xi) = \tfrac{C_1}{a} \xi +\tfrac{C_2}{a} + C_3 \cos(\sqrt{\tfrac{a}{b}}\xi +C_4)
\end{equation}
where $C_3,C_4$ are arbitrary constants.
This yields profiles $V(\xi)$ that are non-symmetric about their peak when $C_1\neq0$. 
The analogous reduction for the $K(m,n)$ equation,
which has $\k =\nu$, holds only for $\nu=0$,
namely when the compacton is static.

\subsection{Multi-reduction}

The reductions \eqref{V.quadrature} and \eqref{V.lin}
can be derived in an alternative way by symmetry considerations,
starting from conservation laws of the $\KP(m,n)$ equation \eqref{KPmn}.

A local conservation law is a continuity equation
$D_t T + D_x X + D_yY =0$ holding on all solutions $u(t,x,y)$ of equation \eqref{KPmn},
where $T$ is the conserved density and $(X,Y)$ is the flux.
(See \Ref{Olv,BCA,Anc-review} for a review of conservation laws for PDEs.)
Any conservation law that is invariant under the two translation symmetries
$\X_1 = \nu \partial_x +\partial_t$ and $\X_2 = \mu \partial_x -\partial_y$
will reduce to a first integral of the travelling wave ODE \eqref{U.ODE},
given by $(X+\mu Y-\nu T)'=0$ for all travelling wave solutions $u=U(\xi)$.

The method of multi-reduction introduced in \Ref{AncGan2020}
can be applied to find all symmetry-invariant conservation laws,
including ones that are admitted only for special cases of $m\neq1$, $n$, $\mu$, $\nu$.
In the general case,
two conservation laws are obtained:
\begin{equation}\label{conslaw1}
\begin{aligned}
T_1 = & 0 ,
\\
X_1 = & u_t + a mu^{m-1}u_x
+bn\big( (n-1)(n-2)u^{n-3}u_x^3+3(n-1)u^{n-2}u_x u_{xx}+u^{n-1}u_{xxx} \big),
\\
Y_1=& \s u_y ;
\end{aligned}
\end{equation}
and
\begin{equation}\label{conslaw2}
\begin{aligned}
T_2 = & -\tfrac{1}{2} u,
\\
X_2 = & -au^m +(x+\mu y-\nu t)\big( u_t +a m u^{m-1}u_x
+ b n\big( (n-1)(n-2) u^{n-3}u_x^3
\\ &
+3 (n-1) u^{n-2}u_x u_{xx} +(n-1)u^{n-1} u_{xxx} \big) \big)
- bn\big( (n-1) u^{n-2}u_x^2 + u^{n-1} u_{xx} \big),
\\
Y_2= & \s\mu u + \s(x+\mu y-\nu t)u_y .
\end{aligned}
\end{equation}
These conservation laws respectively represent 
a topological charge \cite{AncRec} and a mass. 

The only special case admitting additional symmetry-invariant conservation laws is
$n=m\neq 1$, $\nu=\s^2\mu$ ($\k=0$). 
The admitted conservation laws are given by
\begin{equation}\label{conslaw3}
\begin{aligned}
T_3 = & \sqrt{\tfrac{a}{b}} \cos\Big( \sqrt{\tfrac{a}{b}}(x+\mu y -\s \mu^2 t) \Big)u,
\\
X_3 = &
n\sqrt{ab} \cos\Big( \sqrt{\tfrac{a}{b}}(x+\mu y -\s \mu^2 t) \Big)
\big( (n-1) u^{n-2}u_x^2 +u^{n-1}u_{xx} \big)
\\&
+ \sin\Big( \sqrt{\tfrac{a}{b}}(x+\mu y -\s \mu^2 t) \Big)
\big( u_t +bn\big( (n-1)(n-2) u^{n-3}u_x^3
\\&
+(n-1) u^{n-2}u_x u_{xx} + u^{n-1}u_{xxx} \big) \big),
\\
Y_3= & \s\mu \sqrt{\tfrac{a}{b}} \cos\Big( \sqrt{\tfrac{a}{b}}(x+\mu y -\s\mu^2 t) \Big)u
+ \s \sin\Big( \sqrt{\tfrac{a}{b}}(x+\mu y -\s\mu^2 t) \Big)  u_y ,
\end{aligned}
\end{equation}
and a similar conservation law with cosine and sine being permuted.
Each of these additional conservation laws represent a weighted mass. 

Combining the first integrals that arise from the conservation laws \eqref{conslaw1} and \eqref{conslaw2}
yields the reduced second-order ODE \eqref{V.reducedODE}.
The two conservation laws admitted in the special case
yield two additional first integrals which combine with previous two first integrals
to give the explicit expression \eqref{V.lin}.

\subsection{Solitary waves}

A \emph{line solitary wave} is a travelling wave \eqref{travellingwave}
whose profile $U(\xi)$ asymptotically goes to zero as $|\xi|\to 0$.
This implies $C_1=C_2=0$ in the ODE \eqref{U.reducedODE}
which then can be integrated to get
\begin{equation}\label{U.ODE.solitary}
U'{}^2 +A U^{2+m-n} -B U^{3-n} =0,
\quad
A =  \tfrac{2a}{n(m+n)b}, 
\quad
B = \tfrac{2\k}{n(n+1)b},
\end{equation}
where the integration constant is zero due to the asymptotic condition on $U(\xi)$.

This first-order ODE has the form of a nonlinear oscillator equation
$U'{}^2 +\V(U)=0$
with the potential being 
\begin{equation}\label{solitary.potential}
\V(U) = A U^{2+m-n} -B U^{3-n}
\end{equation}
and the energy being $0$. 
Existence of a solitary wave requires that the powers $3-n$ and $2+m-n$ are positive
such that the potential has 
a non-zero root $U=U_*$
and a local maximum or an inflection at $U=0$.
The solitary wave will have a bright or dark peak $U=U_*$,
and a tail $U\to 0$ which will have 
exponential decay in $\xi$ in the case of a local maximum, 
$\V(U)/U^2 \sim \V''(0)<0$,
or power decay in $\xi$ in the case of an inflection, 
$\V(U)/U^2 \sim 0$. 
These cases respectively describe
a localized solitary wave 
and a heavy-tail solitary wave.
The latter type of solution is known to occur in the gKdV equation $K(m,1)$
when solitary waves on a non-zero background are considered 
(see \Ref{AncNayRec} and references therein).

Analysis of the conditions for existence of a solitary wave is straightforward
by considering the two cases $m\gtrless1$.

\begin{figure}[h]
\centering
\includegraphics[trim=2cm 17cm 7cm 2cm,clip,width=0.6\textwidth]{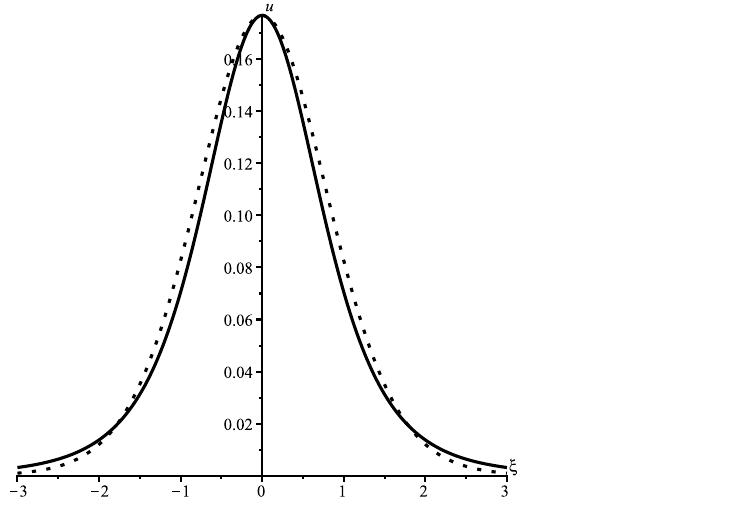}
\caption{Profiles of localized solitary wave \eqref{mlarger1.solitary} (dotted)
and heavy-tail wave \eqref{mlarger1.heavytail} (solid):
$m=\tfrac{7}{5}$, $A=16$, $B=8$.
}\label{fig-mlarger1-solitary}
\end{figure}

When $m>1$, 
the potential \eqref{solitary.potential} has a non-zero root 
iff $\sgn(A)=\sgn(B)$:
$U_*=(B/A)^{1/(m-1)}$.
Then $U=0$ is a local maximum iff $n=1$, or an inflection iff $n<1$,
with $B>0$ in both cases. 
For $n=1$,
the solution of the solitary wave ODE \eqref{U.ODE.solitary} is the well-known \cite{AncGanRec}
line soliton of the generalized KP ($\KP(m,1)$) equation:
\begin{equation}\label{mlarger1.solitary}
U(\xi)=
(\tfrac{B}{A})^{1/(m-1)}\sech(\tfrac{(m-1)\sqrt{B}}{2}\xi)^{2/(m-1)},
\quad
A,B>0,
\quad
n=1,
\quad
m>1 .
\end{equation}
For $n<1$,
the solution of the solitary wave ODE \eqref{U.ODE.solitary} is given by hypergeometric functions.
An explicit solution arises in the special case $n=2-m>0$:
\begin{equation}\label{mlarger1.heavytail}
U(\xi)=
1/\big( \tfrac{A}{B}+\tfrac{(n-1)^2B}{4n^2}\xi^2 \big)^{1/(1-n)},
\quad
A,B>0;
\quad
n=2-m,
\quad
1<m<2 .
\end{equation}
This is a heavy-tail wave, which decays to zero as a power $|\xi|^{-1/(m-1)}$.
Note that the nonlinear dispersion power is sublinear, $n<1$.
A plot comparing these two solutions is shown in Fig.~\ref{fig-mlarger1-solitary}.

\begin{figure}[h]
\centering
\includegraphics[trim=2cm 17cm 6cm 2cm,clip,width=0.6\textwidth]{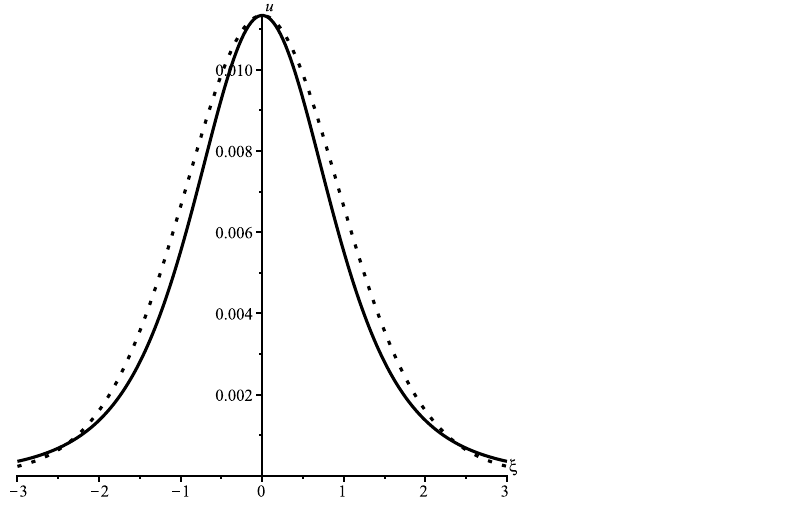}
\caption{Profiles of solitary wave \eqref{msmaller1.solitary} (dotted)
and heavy-tail wave \eqref{msmaller1.heavytail} (solid):
$m=\tfrac{3}{5}$, $A=-2$, $B=-12$.
}\label{fig-msmaller1-solitary}
\end{figure}

When $m<1$, 
again the potential \eqref{solitary.potential} has a non-zero root 
iff $\sgn(A)=\sgn(B)$:
$U_*=(\tfrac{A}{B})^{1/(1-m)}$.
Then $U=0$ is a local maximum iff $n=m$, or an inflection iff $n<m$,
with $A,B<0$.
In the case $n=m$,
the solution to the solitary wave ODE \eqref{U.ODE.solitary} is explicitly given by
\begin{equation}\label{msmaller1.solitary}
U(\xi) = (|\tfrac{A}{B}|)^{1/(1-m)} \sech\big( \tfrac{(1-m)\sqrt{|A|}}{2m} \xi \big)^{2/(1-m)},
\quad
n=m,
\quad
m<1 .
\end{equation}
This is a standard sech type profile.
In the case $n<m$,
the solution is given by hypergeometric functions.
An explicit solution arises in the special case $n=2m-1>0$:
\begin{equation}\label{msmaller1.heavytail}
U(\xi)=
1/\big( |\tfrac{B}{A}|+\tfrac{(n-1)|A|}{4n^2} \xi^2 \big)^{2/(1-n)},
\quad
n=2m-1,
\quad
\tfrac{1}{2}<m<1 .
\end{equation}
This is a heavy-tail wave, which decays to zero as a power $|\xi|^{-1/(1-m)}$.
Note that the nonlinearity powers here are sublinear.
A plot comparing these two solutions is shown in Fig.~\ref{fig-msmaller1-solitary}

The solutions \eqref{mlarger1.heavytail}, \eqref{msmaller1.solitary}, \eqref{msmaller1.heavytail}
have not appeared previously in the literature.

\section{Conditions for existence of compacton solutions}\label{sec:conditions}

Conditions will now be derived on the form of a travelling wave solution $U(\xi)$, 
with a peak at $\xi=0$, 
so that it determines a compacton solution given by
\begin{equation}\label{cutoff.soln}
U_\c(\xi) = U(\xi)H(L-|\xi|)
\end{equation}
where $H(\xi)$ is the Heaviside step function.
Namely, up to an irrelevant translation in $\xi$, 
the travelling wave profile is cut off at $|\xi|=L$ to obtain the compacton,
which has width $2L$ and is symmetric about $\xi=0$. 

A wave profile of the form \eqref{cutoff.soln} will be a classical (strong) solution of
the fourth-order ODE \eqref{U.ODE}
only if $U_\c(\xi)$ is $C^4$ on $-\infty< \xi< \infty$. 
At the cut-off $\xi=\pm L$, this requires 
\begin{equation}\label{strong.conds}
U(\pm L)=0,
\quad
U'(\pm L)=0,
\quad
U''(\pm L)=0,
\quad
U'''(\pm L)=0,
\quad
U''''(\pm L)=0.
\end{equation}

\begin{rem}
Note that if some terms in the ODE \eqref{U.ODE} contain negative powers of $U$
then those terms may be singular due to $U$ vanishing at $\xi =\pm L$ 
or $U$ changing sign at some other point $\xi=\xi_0$. 
In such situations, 
the ODE can be recast by multiplying it by a suitable positive power of $U$ 
so that all of the terms are non-singular.
For arbitrary $n,m>0$,  it is sufficient to multiply by $U^s$ with $s=\max(4-n,2-m,0)$:
\begin{equation}\label{U.ODE.recast}
b (U^n)'''' U^s +a (U^m)'' U^s  -\k U'' U^s =0 .
\end{equation}
This will be assumed hereafter whenever necessary. 
\end{rem}

To investigate conditions \eqref{strong.conds}, 
suppose that $U$ near the cut off has an asymptotic behaviour 
\begin{equation}\label{U.asympt}
U(\xi) \sim c_\pm\, (L\mp \xi)^p  \text{ as } \xi \sim \pm L
\end{equation}
for some power $p$, where $c^+$ and $c^-$ are non-zero constants.
Note that the corresponding asymptotic behaviour on $V(\xi)$ is given by
\begin{equation}\label{V.asympt}
V(\xi) \sim c_\pm^n\, (L\mp \xi)^{pn}  \text{ as } \xi \sim \pm L . 
\end{equation}
Substitution of expression \eqref{U.asympt}
into the necessary and sufficient conditions \eqref{strong.conds}
immediately yields the following result.

\begin{prop}\label{prop:p.conds}
For a travelling wave $U(\xi)$ with asymptotic behaviour \eqref{U.asympt}, 
a compacton profile \eqref{cutoff.soln} gives 
a classical (strong) solution of the travelling wave ODE \eqref{U.ODE.recast}
iff the cut-off power satisfies 
\begin{equation}\label{strong.p.conds}
p>4.
\end{equation}
\end{prop}

The power $p$ that arises for a given travelling wave solution $U(\xi)$ is determined by
the quadrature \eqref{V.quadrature} in the general case $m,n>0$,
and by the explicit expression \eqref{V.lin} in the special case $m=n>0$, $\k=0$. 
Once $p$ is found,
the condition \eqref{strong.p.conds} can be reduced to inequalities on $(n,m)$,
which will be necessary and sufficient 
for the compacton profile \eqref{cutoff.soln} to be 
a classical solution. 
This analysis will be carried out 
for the general case \eqref{V.quadrature} and the special case \eqref{V.lin}
respectively in the next two sections.
The results will be seen to yield a wide variety of compacton solutions.

It is interesting to observe that there is a more general notion of a solution
which arises from considering the travelling wave ODE \eqref{U.ODE}
in the twice-integrated form \eqref{U.reducedODE}
when the two integration constants are put to zero:
\begin{equation}\label{U.ODE.2ndord}
b(U^n)'' + aU^m -\k U=0 . 
\end{equation}
For arbitrary $n,m>0$,  
multiplication of this ODE by $U^s$ with $s=\max(2-n,0)$
ensures that all of the terms are non-singular, 
\begin{equation}\label{U.ODE.2ndord.recast}
b(U^n)''U^s + aU^{m+s} -\k U^{1+s} =0 ,
\end{equation}
so that compactons will exist as classical solutions. 
Then the counterpart of Proposition~\ref{prop:p.conds} is given by the following result.

\begin{prop}\label{prop:p.conds.2ndord}
For a travelling wave $U(\xi)$ with the asymptotic behaviour \eqref{U.asympt}, 
the compacton profile \eqref{cutoff.soln} yields
a classical (strong) solution of the second-order ODE \eqref{U.ODE.2ndord.recast}
iff the cut-off power satisfies the condition
\begin{equation}\label{strong.p.conds.2ndord}
p>2.
\end{equation}
\end{prop}

There turn out to be compacton profiles that satisfy this condition \eqref{strong.p.conds.2ndord}
but not the stronger condition in Proposition~\ref{prop:p.conds}.  
However, while such profiles are solutions of the twice-integrated ODE, 
they {\em are not} classical solutions \eqref{travellingwave} of
the travelling wave ODE \eqref{U.ODE}
and thus do not yield solutions of the $\KP(m,n)$ equation \eqref{KPmn}.

\section{Compactons in a special case}\label{sec:linearcase}

All line compactons in the case $m=n$, $\k=0$ 
will now be derived, 
starting from the travelling wave solutions given by expression \eqref{V.lin} in terms of $V(\xi)$.
Since this expression is an analytic function of $\xi$ that involves four constants $C_1,C_2,C_3,C_4$,
all solutions $V(\xi)$ have the asymptotic behaviour \eqref{V.asympt}
with $p$ being determined by the values of those four constants and the value of $L>0$.

From Proposition~\ref{prop:p.conds},
the necessary condition $p>0$ requires that $V(\pm L)=0$.
This gives two equations
\begin{equation}\label{V.lin.L}
\pm C_1 L + C_2 + C_3 a\cos(\sqrt{\tfrac{a}{b}}L \mp C_4) =0 ,
\end{equation}
which can be expanded and combined to get 
\begin{equation}\label{V.lin.C1C2}
C_1 = -(a/L) C_3 \sin(C_4)\sin(\sqrt{\tfrac{a}{b}}L) ,
\quad
C_2  = -a C_3 \cos(C_4) \cos(\sqrt{\tfrac{a}{b}}L) .
\end{equation}
Hence, expression \eqref{V.lin} becomes
\begin{equation}\label{V.lin.compacton}
V(\xi) = C_3 \big(
\cos(\sqrt{\tfrac{a}{b}}\xi + C_4)
+\sin(\sqrt{\tfrac{a}{b}}L)\sin(C_4) (\xi/L)
-\cos(\sqrt{\tfrac{a}{b}}L)\cos(C_4)
\big) ,
\end{equation}
which contains three constants $C_3\neq 0$, $C_4$, and $L>0$.
Note that, because $C_3$ appears as an amplitude factor,
the power $p$ in the asymptotic behaviour \eqref{V.asympt}
is determined just by the values of $C_4$ and $L$.

For arbitrary $C_4$ and $L>0$,
expression \eqref{V.lin.compacton} has $pn=1$.
It thereby yields a classical compacton solution if and only if $p=\frac{1}{n}>4$, 
as shown by Proposition~\ref{prop:p.conds}. 
Likewise, 
it is also a compacton solution of the twice-integrated ODE \eqref{U.ODE.2ndord}
if and only if $p=\frac{1}{n}>2$, 
as a consequence of Proposition~\ref{prop:p.conds.2ndord}. 

Furthermore, expression \eqref{V.lin.compacton} can have $pn=2$
if $C_4$ and $L$ are chosen such that $V'(\pm L)=0$.
This condition yields the equations
\begin{equation}
\sin(\sqrt{\tfrac{a}{b}L})\cos(C_4) =0,
\quad
\sin(C_4)\big( \sqrt{\tfrac{a}{b}}L \cos(\sqrt{\tfrac{a}{b}}L) -\sin(\sqrt{\tfrac{a}{b}}L) \big)=0 .
\end{equation}
There are three solutions:
\begin{align}
&
C_4 = 0,
\quad
L = \sqrt{\tfrac{b}{a}} \pi ;
\\
&
C_4 = 0,
\quad
L = \sqrt{\tfrac{b}{a}} 2\pi ;
\\
&
C_4 = \pm\tfrac{1}{2}\pi,
\quad
L = \sqrt{\tfrac{b}{a}} z,
\quad
z=\tan(z),
\quad
z>0 .
\end{align}
Hence, expression \eqref{V.lin.compacton} becomes, respectively,
\begin{align}
& V(\xi) = 2 C_3 \cos(\tfrac{1}{2}\sqrt{\tfrac{a}{b}}\xi)^2 ,
\label{V.lin.compacton1.pis2}
\\
& V(\xi) = -2 C_3 \sin(\tfrac{1}{2}\sqrt{\tfrac{a}{b}}\xi)^2 ,
\label{V.lin.compacton2.pis2}
\\
&
V(\xi) = C_3 \big( \sqrt{\tfrac{a}{b}} \cos(\sqrt{\tfrac{a}{b}}L) (\xi/L) -\sin(\sqrt{\tfrac{a}{b}}\xi)  \big) .
\label{V.lin.compacton3.pis2}
\end{align}

Since expressions \eqref{V.lin.compacton1.pis2}--\eqref{V.lin.compacton3.pis2}
contain no arbitrary constants other than $C_3$,
there are no special cases in which they satisfy $V''(\pm L)=0$.

Therefore, altogether four classes of compacton solutions are obtained.

\begin{thm}\label{thm:compactons.lin}
(i) 
For the travelling wave ODE \eqref{U.ODE} with $n=m>0$ (where $m\neq1$), $\k=0$,
all compacton classical (strong) solutions \eqref{cutoff.soln} 
are given by
\begin{equation}\label{U.Vlin.strong}
U = \alpha \Big( 
\cos(\sqrt{\tfrac{a}{b}}\xi + \phi)
+\sin(\sqrt{\tfrac{a}{b}}L)\sin(\phi) (\xi/L)
-\cos(\sqrt{\tfrac{a}{b}}L)\cos(\phi)
\Big)^{1/n} ,
\quad
L>0
\end{equation}
with $n<\tfrac{1}{4}$;
\begin{align}
&\begin{aligned}
U =
\alpha \Big( \cos(\tfrac{1}{2}\sqrt{\tfrac{a}{b}}\xi) \Big)^{2/n},
\quad
L = \sqrt{\tfrac{b}{a}} \pi ,
\end{aligned}\label{U.Vlin1.weak}
\\
&\begin{aligned}
U =
\alpha \Big( \sin(\tfrac{1}{2}\sqrt{\tfrac{a}{b}}\xi) \Big)^{2/n},
\quad
L = \sqrt{\tfrac{b}{a}} 2\pi ,
\end{aligned}\label{U.Vlin2.weak}
\\
&\begin{aligned}
U =
\alpha \Big( \cos(\sqrt{\tfrac{a}{b}}L) \sqrt{\tfrac{a}{b}} \xi  -\sin(\sqrt{\tfrac{a}{b}}\xi)  \Big)^{1/n},
\quad
L = \sqrt{\tfrac{b}{a}} z,\
z=\tan(z), \
z>0 ,
\end{aligned}\label{U.Vlin3.weak}
\end{align}
with $n<\frac{1}{2}$. 
Here 
\begin{equation}
\xi = x + \mu y - \s\mu^2 t
\end{equation}
is the travelling wave variable; 
$\alpha$ is an arbitrary positive constant, and $\phi$ is an arbitrary constant mod $\pi$.
\newline
(ii)
For the twice-integrated ODE \eqref{U.ODE.2ndord}
with $m=n>0$ (where $m\neq1$), $k=0$,
all compacton solutions \eqref{cutoff.soln} are given by 
expression \eqref{U.Vlin.strong} with $n<\frac{1}{2}$,
and expressions \eqref{U.Vlin1.weak}, \eqref{U.Vlin2.weak}, \eqref{U.Vlin3.weak}
with $n<1$.
\end{thm}

The solutions \eqref{U.Vlin.strong} and \eqref{U.Vlin1.weak}--\eqref{U.Vlin3.weak}
describe line compactons $u=U_\c(\xi)$ of the $K_2(m,n)$ equation \eqref{KPmn}.

\subsection{Kinematic features and profiles}
These compactons have direction $\theta=\arctan(\mu)$ 
and speed $c= \s\mu^2/\sqrt{1+\mu^2}$ 
determined by the parameter $\mu$.
Thus, $c$ and $\theta$ are constrained by the kinematic relation
\begin{equation}\label{speed.dir}
c = \s \sin(\theta)^2/|\cos(\theta)| .
\end{equation}
With respect to the $x$ axis,
faster waves move more transversely,
while slower waves move more tangentially. 

There are three different types of profiles.

\begin{figure}[h]
\centering
\includegraphics[trim=2cm 15cm 5cm 6cm,clip,width=0.45\textwidth]{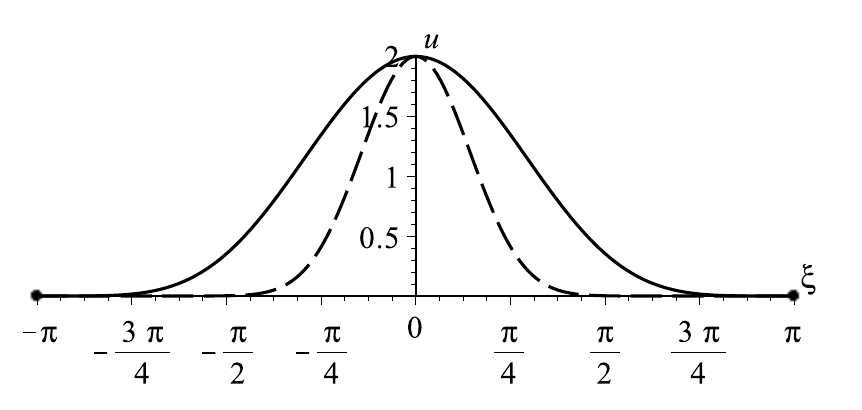}
\caption{Cosine profiles \eqref{U.Vlin1.weak} with $\frac{a}{b}=1$:
$n=\tfrac{4}{9}$ solid, $n=\tfrac{1}{10}$ dash.
Dots indicate the cut off points.}\label{fig-cos}
\end{figure}

A cosine profile is exhibited by the compactons \eqref{U.Vlin1.weak}.
This profile is non-negative and symmetric in $\xi$.
Plots are shown in Fig.~\ref{fig-cos}.

\begin{figure}[h]
\centering
\includegraphics[trim=2cm 15cm 5cm 6cm,clip,width=0.45\textwidth]{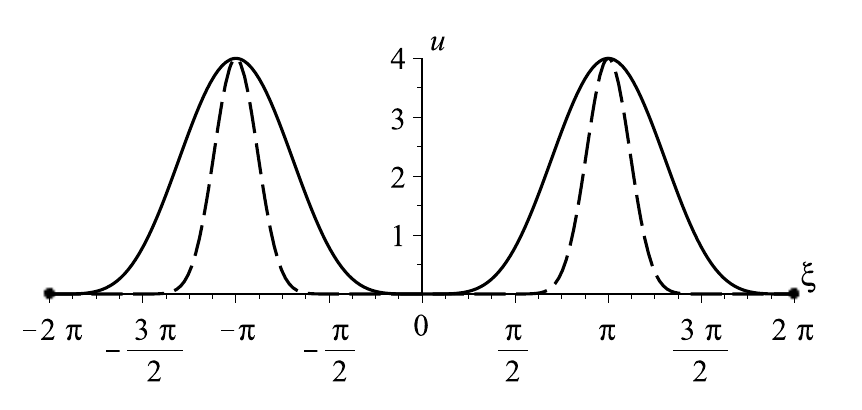}
\includegraphics[trim=2cm 14cm 5cm 4cm,clip,width=0.45\textwidth]{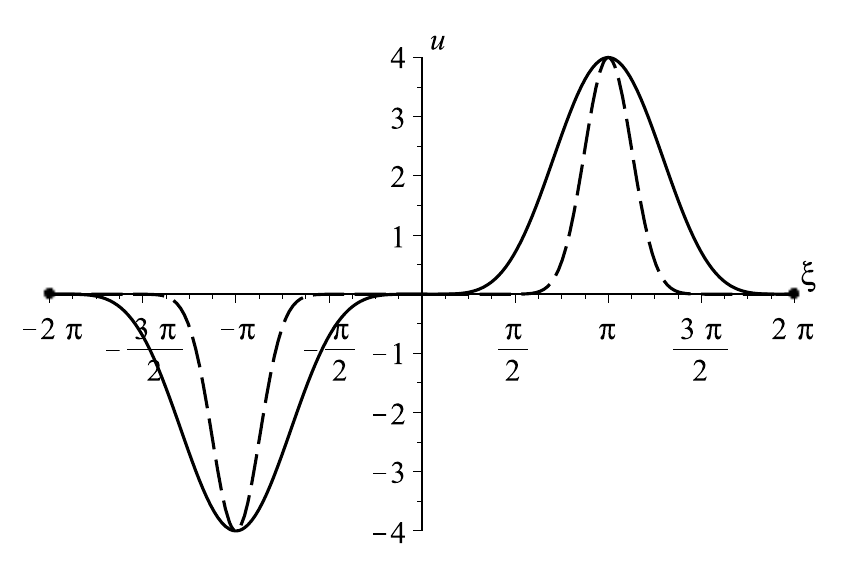}
\caption{Sine profiles \eqref{U.Vlin2.weak} with $\frac{a}{b}=1$.
Left: $n=\tfrac{2}{5}$ solid, $n=\tfrac{2}{25}$ dash;
Right: $n=\tfrac{3}{7}$ solid, $n=\tfrac{1}{15}$ dash.
Dots indicate the cut off points.}\label{fig-sin}
\end{figure}

A sine profile is exhibited by the compactons \eqref{U.Vlin2.weak}.
For this profile to be well-defined for both negative and positive parts of the sine,
a restriction is necessary on $n$ in the power $\tfrac{2}{n}$. 
If this power is a rational number whose numerator is even,
the resulting sine profile is effectively squared so that 
the compacton is non-negative and symmetric in $\xi$ with a node at $\xi=0$.
Instead if the numerator and denominator are odd,
the compacton is antisymmetric (sign-changing) in $\xi$.
Plots are shown in Fig.~\ref{fig-sin}

\begin{figure}[h]
\centering
\includegraphics[trim=2cm 19cm 12cm 1cm,clip,width=0.42\textwidth]{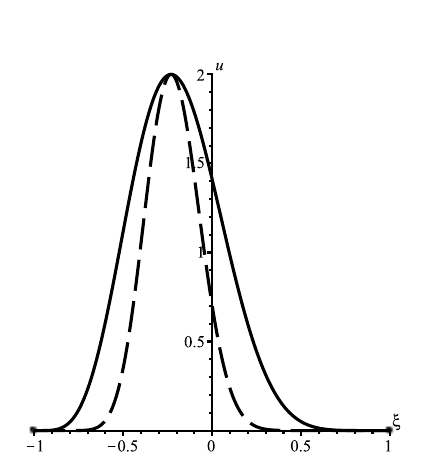}
\includegraphics[trim=2cm 19cm 12cm 1cm,clip,width=0.45\textwidth]{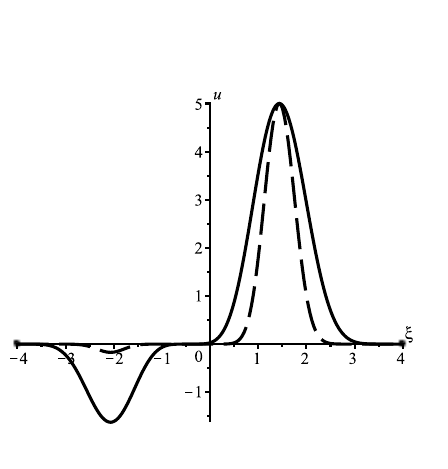}
\caption{Non-symmetric profiles \eqref{U.Vlin.strong} with $\frac{a}{b}=1$:
$n=\tfrac{1}{5}$ solid, $n=\tfrac{1}{10}$ dash. 
Left: $\phi=\tfrac{\pi}{3}$, $L=1$;
Right: $\phi=\tfrac{2\pi}{5}$, $L=4$.
Dots indicate the cut off points.}\label{fig-nonsymm2}
\end{figure}

In contrast,
a novel non-symmetry profile is exhibited by the compactons \eqref{U.Vlin.strong}
and \eqref{U.Vlin3.weak}. 
Depending on the parameter values, the profile can be strictly non-negative or non-positive, or sign-changing with a node at some $\xi=\xi_0$.
Hence, a restriction on $n$ similar to the sine case must hold in the latter two cases.
See plots in Fig.~\ref{fig-nonsymm2}.

\begin{rem}\label{rem:misn.kis0.static}
The preceding line compactons have a reduction to the $K(m,n)$ equation 
by setting $\mu=0$ and $\s=0$. 
Then the kinematic relation \eqref{speed.dir} shows that $c=\nu=0$,
whereby the resulting compactons are static. 
\end{rem}

\section{Compactons in the general case}\label{sec:nonlinearcase}

All line compactons arising from the quadrature \eqref{V.quadrature}
for travelling wave solutions in terms of $V(\xi)$
will be derived next. 

Since both Propositions~\ref{prop:p.conds} and~\ref{prop:p.conds.2ndord}
require that $V(\pm L)=0$,
the quadrature for relevant solutions $V(\xi)$ can be expressed more conveniently as
\begin{equation}\label{V.quadrature.compacton}
\int_0^V \frac{dV}{\sqrt{ E +C V + B V^{1+1/n} -A V^{1+m/n} }}
= L \mp\xi 
\end{equation}
where
\begin{gather}
L = C_4>0,
\quad
E=\tfrac{2C_3}{b},
\quad
C = \tfrac{2C_2}{b},
\\
B = \tfrac{2n\k}{(n+1)b},
\quad
A =  \tfrac{2na}{(m+n)b},
\quad
\k = \nu- \s\mu^2
\label{B.A.kappa}
\end{gather}
are constants.
This form of the quadrature assumes $V\geq 0$, 
and the required integration domain is thereby $0\leq V\leq V_{\max}$
where $V_{\max}$ is the smallest positive non-repeated root of the denominator. 
In particular, $V_{\max}$ must satisfy
\begin{equation}\label{Vmax}
V_{\max}>0,
\quad
E +C V_{\max} + B V_{\max}^{1+1/n} -A V_{\max}^{1+m/n} =0 ,
\quad
C + (1+\tfrac{1}{n})B V_{\max}^{1/n} -(1+\tfrac{m}{n})A V_{\max}^{m/n} \neq 0  .
\end{equation}

\begin{rem}
Solutions with $V\leq0$ can be obtained by a suitably altering the quadrature.
More generally, solutions in which $V$ changes sign can be found. 
Hereafter, only the case $V\geq0$ will be considered for simplicity. 
\end{rem}

Near $\xi=\pm L$, the asymptotic behaviour of $V(\xi)$ is determined by
the term that has the smallest exponent appearing 
in the denominator of the quadrature \eqref{V.quadrature.compacton}:
$E$, $C V$, $B V^{1+1/n}$, $A V^{1+m/n}$,
where $m,n>0$ and $m\neq1$. 
The subsequent analysis, based on Proposition~\ref{prop:p.conds},
will be split into cases depending on whether each coefficient $E,C,B$ is non-zero or zero.
Note that, due to $a\neq 0$, there is no case in which $A=0$. 
\newline

\emph{Case $E\neq 0$}:
The smallest exponent in this case is given by the constant term $E$.
Then the quadrature \eqref{V.quadrature.compacton} yields
$V \sim \sqrt{2E}\,(L \mp \xi)$ as $\xi\sim \pm L$.
This requires $E>0$.
Since the resulting asymptotic behaviour \eqref{V.asympt} has $pn=1$,
a compacton exists for $p=\frac{1}{n}>4$
according to condition \eqref{strong.p.conds}. 
This gives $n<\tfrac{1}{4}$. 

The additional condition \eqref{Vmax} requires that at least one of 
$B\neq0$ and $C\neq0$ holds. 
It is convenient to split up the analysis into cases $B=0$ and $B\neq 0$, 
and for the latter, make a further splitting into subcases $m\gtrless1$, 
and then finer subcases with respect to $\sgn(C)$. 
Consider $E = P(V)>0$ where 
\begin{equation}\label{Epos.P}
P(V)=A V^{1+m/n} -B V^{1+1/n} -C V .
\end{equation}

\emph{Subcase $B=0$, $C\neq0$}.
Clearly, there exists a positive root $V_{\max}$ of $P(V)=E$ if $A>0$, 
since $P(V)$ vanishes for $V=0$ and exceeds $E>0$ for large enough $V>0$.
If $A<0$, existence of a positive root requires that $P(V)-E$ has a non-negative maximum for some $V=V_0>0$. 
The only extremum of $P(V)$ is $V_0= \big(\tfrac{n|C|}{(n+m)|A|}\big)^{n/m}$ with $C<0$. 
This gives the condition 
\begin{equation}
E\leq \tfrac{m}{n+m}|C| V_0 .
\end{equation}

\emph{Subcase $B\neq0$, $m>1$}.
In this case, note that the term in $P(V)$ having the largest exponent is $AV^{1+m/n}$. 
There exists a positive root $V_{\max}$ if $A>0$, 
since $P(V)$ vanishes for $V=0$ and exceeds $E>0$ for large enough $V>0$.

If $A<0$, existence of a positive root requires that $P(V)-E$ has a non-negative maximum
for some $V=V_0>0$. 
The condition for an extremum can be expressed as $R(V_0)=C$ where 
\begin{equation}\label{Epos.mgtr1.R}
R(V) = -(1+\tfrac{m}{n})|A| V^{m/n} - (1+\tfrac{1}{n})B V^{1/n} ,
\end{equation}
with the term $V^{m/n}$ having the largest exponent. 

If $C=0$, then $R(V_0)=0$ yields 
$V_0 = \big(\tfrac{(n+1)|B|}{(n+m)|A|}\big)^{n/(m-1)}>0$ 
provided that $B<0$. 
The condition $P(V_0)-E\geq 0$ combined with $E>0$ gives the inequality
\begin{equation}
0< E\leq P(V_0) = \tfrac{(m-1)|B|}{m+n} V_0^{1+1/n} . 
\end{equation}

If $C<0$, then a root $V_0>0$ of $R(V_0)=-|C|$ clearly exists, 
since $-R(V)$ vanishes for $V=0$ and exceeds $|C|$ for large enough $V>0$ 
due to $(1+\tfrac{m}{n})|A| V^{m/n}$ being the leading term. 
This extremum must satisfy 
\begin{equation}
0<E\leq P(V_0) = \tfrac{1}{n+1}|C| V_0 +\tfrac{m-1}{n+1}|A| V_0^{1+m/n} .
\end{equation}

If $C>0$, existence of a root $V_0>0$ of $R(V_0)=C$ 
can be seen graphically to require $B<0$ such that 
$R'(V_1)=0$ and $R(V_1)>C$ for some $0<V_1<V_0$. 
The only extremum of $R'(V)$ is 
$V_1= \big(\tfrac{(n+1)|B|}{m(m+n)|A|}\big)^{n/(m-1)}>0$, 
which yields the inequality
\begin{equation}
R(V_1)=\tfrac{(n+1)(m-1)|B|}{nm} V_1^{1/m} >C .
\end{equation}
Then the root $V_0>0$ given by 
\begin{equation}
R(V_0) = (1+\tfrac{1}{n})|B| V_0^{1/n} -(1+\tfrac{m}{n})|A| V_0^{m/n} =C 
\end{equation} 
must satisfy the conditions
\begin{equation}
0<E\leq P(V_0) = \tfrac{m}{n}|A| V_0^{1+m/n} -\tfrac{1}{n}|B| V_0^{1+1/n}
\end{equation}
and
\begin{equation}
V_0>\big(\tfrac{(n+1)|B|}{m(m+n)|A|}\big)^{n/(m-1)}
\end{equation}

\emph{Subcase $B\neq0$, $m<1$}.
In this case, the term in $P(V)$ having the largest exponent is $BV^{1+1/n}$. 
A positive root $V_{\max}$ exists if $B<0$, 
since then $P(V)$ vanishes for $V=0$ and exceeds $E>0$ for large enough $V>0$.

If $B>0$, existence of a positive root requires that $P(V)-E$ has a non-negative maximum
for some $V=V_0>0$. 
The condition for an extremum can be expressed as $R(V_0)=C$ where 
\begin{equation}\label{Epos.mlss1.R}
R(V) = (1+\tfrac{m}{n})A V^{m/n} -(1+\tfrac{1}{n})B V^{1/n} , 
\end{equation}
with the term $V^{1/n}$ having the largest exponent. 

If $C=0$, then $R(V_0)=0$ yields 
$V_0 = \big(\tfrac{(n+1)B}{(n+m)A}\big)^{n/(m-1)}>0$ 
provided that $A>0$. 
Combining $E>0$ with the condition $P(V_0)-E\geq 0$ 
gives the inequality
\begin{equation}
0< E\leq P(V_0) = \tfrac{(1-m)B}{m+n} V_0^{1+1/n} . 
\end{equation}

If $C<0$, then a root $V_0>0$ of $R(V_0)=-|C|$ clearly exists, 
since $-R(V)$ vanishes for $V=0$ and exceeds $|C|$ for large enough $V>0$ 
due to $(1+\tfrac{1}{n})B V^{1/n}$ being the leading term. 
This extremum must satisfy 
\begin{equation}
0<E\leq P(V_0) = \tfrac{1}{n+1}|C| V_0 +\tfrac{1-m}{n+1}A V_0^{1+m/n} .
\end{equation}

If $C>0$, existence of a root $V_0>0$ of $R(V_0)=C$ 
can be seen graphically to require $A>0$ such that 
$R'(V_1)=0$ and $R(V_1)>C$ for some $0<V_1<V_0$. 
The only extremum of $R'(V)$ is 
$V_1= \big(\tfrac{(n+1)B}{m(m+n)A}\big)^{n/(m-1)}>0$, 
which yields the inequality
\begin{equation}
R(V_1)=\tfrac{(n+1)(1-m)B}{nm} V_1^{1/m} >C .
\end{equation}
Then the root $V_0>0$ given by 
\begin{equation}
R(V_0) = (1+\tfrac{m}{n})A V_0^{m/n} - (1+\tfrac{1}{n})B V_0^{1/n} =C 
\end{equation} 
must satisfy the conditions
\begin{equation}
0<E\leq P(V_0) = \tfrac{1}{n}B V_0^{1+1/n} - \tfrac{m}{n}A V_0^{1+m/n} 
\end{equation}
and
\begin{equation}
V_0>\big(\tfrac{(n+1)B}{m(m+n)A}\big)^{n/(m-1)}
\end{equation}
\newline

\emph{Case $E=0$, $C\neq 0$}:
Among the non-constant terms in the square root 
appearing in the quadrature \eqref{V.quadrature.compacton}, 
$CV$ has the smallest exponent.
Thus, the quadrature yields
$\sqrt{V} \sim \sqrt{C/2}\,(L \mp \xi)$ as $\xi\sim \pm L$.
This requires $C>0$,
and the resulting asymptotic behaviour \eqref{V.asympt} has $pn=2$.
Similarly to the previous case, 
a compacton exists for $p=\frac{2}{n}>4$, which gives $n<\tfrac{1}{2}$. 

The additional condition \eqref{Vmax} 
now reduces to $P(V_{\max}) = C>0$,
where 
\begin{equation}
P(V) = A V^{m/n} -B V^{1/n}
\end{equation}

\emph{Subcase $B\neq0$}. 
A positive root $V_{\max}$ clearly exists if $A>0$ when $m>1$, 
since $P(V)$ vanishes for $V=0$ and exceeds $C>0$ for large enough $V>0$.
Similarly, when $m<1$, a positive root exists if $B<0$. 

For $m>1$ and $A<0$, and for $m<1$ and $B>0$, 
existence of a positive root requires that $R(V)$ has a non-negative maximum
for some $V=V_0>0$.
The only extremum is given by $V_0=\big(\tfrac{B}{mA}\big)^{n/(m-1)}>0$
with $\sgn(A)=\sgn(B)$. 
Then $P(V_0)=\tfrac{m-1}{m}BV_0^{1/n}$ yields
the inequality 
$C<\tfrac{(m-1)B}{m} V_0^{1/n}$ 
in both cases. 

\emph{Subcase $B=0$}. 
A positive root $V_{\max}$ exits iff $A>0$. 
\newline

\emph{Case $E=C=0$, $B,A\neq0$}:
In this case, the terms in the square root 
appearing in the quadrature \eqref{V.quadrature.compacton} 
have exponents $1+1/n$ and $1+m/n$.
This implies two subcases $m\gtrless1$. 

\emph{Subcase $m>1$}. 
The term with the smallest exponent is $B V^{1+1/n}$,
and so the quadrature \eqref{V.quadrature.compacton} yields
$V^{\frac{1}{2}(1-1/n)} \sim \tfrac{1}{2}(1-\tfrac{1}{n})\sqrt{B}\,(L \mp \xi)$ as $\xi\sim \pm L$.
This requires $B>0$ and $n\neq 1$.
The resulting asymptotic behaviour \eqref{V.asympt} has $pn=\tfrac{2n}{n-1}$,
and thus $p=\frac{2}{n-1}$. 

Condition \eqref{strong.p.conds} for existence of a classical compacton 
gives $1<n<\frac{3}{2}$. 
The additional condition \eqref{Vmax}, which determines $V_{\max}>0$, 
is satisfied iff $A>0$.

\emph{Subcase $m<1$}.
The term with the smallest exponent is $A V^{1+m/n}$.
Thus, the quadrature \eqref{V.quadrature.compacton} yields
$V^{\frac{1}{2}(1-m/n)} \sim \tfrac{1}{2}(1-\tfrac{m}{n})\sqrt{-A}\, (L \mp \xi)$ as $\xi\sim \pm L$,
which requires $A<0$.
The asymptotic behaviour \eqref{V.asympt} thereby has $pn=\tfrac{2n}{n-m}$, 
and so $p=\frac{2}{n-m}$. 

From condition \eqref{strong.p.conds}, 
a classical compacton exists for $m<n<m+\frac{1}{2}$. 
The additional condition \eqref{Vmax}, giving $V_{\max}>0$, 
holds iff $B<0$.
\newline

\emph{Case $E=C=B=0$, $A\neq0$}:
In this case, the square root contains only a single term $-A V^{1+1/n}$,
and a positive $V_{\max}$ satisfying condition \eqref{Vmax} does not exist.
Hence, no compactons exist. 
\newline

After all of the preceding cases are merged, 
the following classes of compacton solutions are obtained
from the quadrature \eqref{V.quadrature.compacton} 
through the change of variable \eqref{V.rel}, 
under the conditions stated in Propositions~\ref{prop:p.conds} and~\ref{prop:p.conds.2ndord}.
Details of the proof are straightforward albeit meticulous and will be omitted. 

\begin{thm}\label{thm:compactons.nonlin}
(i) 
For the travelling wave ODE \eqref{U.ODE} with $n,m>0$ and $m\neq1$, 
all compacton classical (strong) solutions \eqref{cutoff.soln} that have $U(\xi)\geq0$ 
are given by the following three families of quadratures.
\newline
(1)
\begin{equation}\label{U.Vnonlin1}
\int_0^{u} \frac{nU^{n-1}\; dU}{\sqrt{E+ C U^n + B U^{n+1} -A U^{n+m} }}
= L \mp \xi,
\end{equation}
with the cases ($E\neq 0$)
\begin{equation}\label{U.Vnonlin1.mgtr1}
n<\tfrac{1}{4} ,
\quad
m>1: 
\end{equation}
\begin{subequations}\label{U.Vnonlin1.mgtr1.cases}
\begin{align}
&
A>0,\
E>0;
\label{U.Vnonlin1.mgtr1.Agtr0}
\\&
\begin{aligned}& 
A<0,\
C<0,\
0<E\leq \tfrac{1}{n+1}((m-1)|A| V_0^{m/n} +|C|)V_0 \\&
\text{where } 
(1+\tfrac{m}{n})|A|V_0^{m/n} + (1+\tfrac{1}{n})BV_0^{1/n} = |C| ; 
\end{aligned}
\label{U.Vnonlin1.mgtr1.Alss0.Clss0}
\\&
A<0,\
B<0,\
C=0,\
0<E\leq (m-1)(\tfrac{1}{n+m}|B|)^{(n+m)/(m-1)}/(\tfrac{1}{n+1}|A|)^{(n+1)/(m-1)} ;
\label{U.Vnonlin1.mgtr1.Alss0.Cis0}
\\&
\begin{aligned}& 
A<0,\
B<0,\
0<C<\tfrac{m-1}{n} (\tfrac{(n+1)}{m}|B|)^{m/(m-1)}/((n+m)|A|)^{1/(m-1)}, \\&
0<E\leq \tfrac{1}{n}(m|A| V_0^{m/n} -|B|V_0^{1/n}),\ 
V_0> (\tfrac{n+1}{m(n+m)}|B|/|A|)^{n/(m-1)} \\&
\text{where } 
(1+\tfrac{1}{n})|B|V_0^{1/n} - (1+\tfrac{m}{n})|A|V_0^{m/n} = C; 
\end{aligned}
\label{U.Vnonlin1.mgtr1.Alss0.Cgtr0}
\end{align}
and 
\end{subequations}
\begin{equation}
n<\tfrac{1}{4} ,
\quad
m<1:
\label{U.Vnonlin1.mlss1}
\end{equation}
\begin{subequations}\label{U.Vnonlin1.mlss1.cases}
\begin{align}
&
B<0,\
E>0;
\label{U.Vnonlin1.mlss1.Blss0}
\\&
\begin{aligned}& 
B>0,\
C<0,\
0<E\leq \tfrac{1}{n+1}((1-m)A V_0^{m/n} +|C|)V_0 \\&
\text{where } 
(1+\tfrac{1}{n})BV_0^{1/n} - (1+\tfrac{m}{n})AV_0^{m/n} = |C| ; 
\end{aligned}
\label{U.Vnonlin1.mlss1.Bgtr0.Clss0}
\\&
\begin{aligned}& 
A>0,\
B>0,\
0<C<\tfrac{m-1}{n} (\tfrac{(n+1)}{m}B)^{m/(m-1)}/((n+m)A)^{1/(m-1)}, \\&
0<E\leq \tfrac{1}{n}(BV_0^{1/n} -m A V_0^{m/n}),\ 
V_0> (\tfrac{n+1}{m(n+m)}B/A)^{n/(m-1)} \\&
\text{where } 
(1+\tfrac{m}{n})A V_0^{m/n} -(1+\tfrac{1}{n})BV_0^{1/n} = C; 
\end{aligned}
\label{U.Vnonlin1.mlss1.Bgtr0.Cgtr0}
\\&
A>0,\
B>0,\
C=0,\
0<E\leq (1-m)(\tfrac{1}{n+m}B)^{(n+m)/(m-1)}/(\tfrac{1}{n+1}A)^{(n+1)/(m-1)} ;
\label{U.Vnonlin1.mlss1.Bgtr0.Cis0}
\\&
A>0,\
B=0,\
E>0;
\label{U.Vnonlin1.mlss1.Bis0.Agtr0}
\\&
A<0,\
B=0,\
C<0,\
0<E\leq m(\tfrac{1}{n+m}|C|)^{1+n/m}/(\tfrac{1}{n}|A|)^{n/m} ;
\label{U.Vnonlin1.mlss1.Bis0.Alss0}
\end{align}
\end{subequations}
(2)
\begin{equation}\label{U.Vnonlin2}
\int_0^{u} \frac{nU^{n-1}\; dU}{\sqrt{ C U^n + B U^{n+1} -A U^{n+m} }}
= L \mp \xi,
\end{equation}
with the cases ($E=0$, $C\neq 0$)
\begin{equation}\label{U.Vnonlin2.mgtr1}
n<\tfrac{1}{2} ,
\quad
m>1: 
\end{equation}
\begin{subequations}\label{U.Vnonlin2.mgtr1.cases}
\begin{align}
&
A>0,\
C>0;
\label{U.Vnonlin2.mgtr1.Agtr0}
\\&
A<0,\
B<0,\
0<C< (m-1)(\tfrac{1}{m}|B|)^{m/(m-1)}/A^{1/{m-1}} ;
\label{U.Vnonlin2.mgtr1.Alss0}
\end{align}
\end{subequations}
and
\begin{equation}\label{U.Vnonlin2.mlss1}
n<\tfrac{1}{2} ,
\quad
m<1: 
\end{equation}
\begin{subequations}\label{U.Vnonlin2.mlss1.cases}
\begin{align}
&
B<0,\
C>0;
\label{U.Vnonlin2.mlss1.Blss0}
\\&
A>0,\
B>0,\
0<C< (1-m)(\tfrac{1}{m}B)^{m/(m-1)}/A^{1/(m-1)} ; 
\label{U.Vnonlin2.mlss1.Bgtr0.Cgtr0}
\\&
A>0,\
B=0,\
C>0; 
\label{U.Vnonlin2.mlss1.Bis0}
\end{align}
\end{subequations}
(3)
\begin{equation}\label{U.Vnonlin3}
\int_0^{u} \frac{nU^{n-1}\; dU}{\sqrt{ B U^{n+1} -A U^{n+m} }}
= L \mp \xi,
\end{equation}
with the cases ($E=C=0$)
\begin{equation}\label{U.Vnonlin3.mgtr1}
1<n<\tfrac{3}{2} ,
\quad
m>1: 
\end{equation}
\begin{align}
& 
A>0,\
B>0;
\label{U.Vnonlin3.mgtr1.Agtr0}
\end{align}
and
\begin{equation}\label{U.Vnonlin3.mlss1}
m<n<m+\tfrac{1}{2} ,
\quad
m< 1: 
\end{equation}
\begin{align}
& 
A<0,\
B<0; 
\label{U.Vnonlin3.mlss1.Blss0}
\end{align}
Here $B$ and $A$ are given by expressions \eqref{B.A.kappa},
$E$ and $C$ are constants,
$L>0$ is an arbitrary positive constant,
and $\xi$ is the travelling wave variable \eqref{travellingwave}.
\newline
(ii) 
For the twice-integrated ODE \eqref{U.ODE.2ndord}, 
all compacton solutions \eqref{cutoff.soln} that have $U(\xi)\geq0$ 
are given by the previous three families of quadratures, 
with the respective conditions on $n$ changed to 
$n<\tfrac{1}{2}$ in cases \eqref{U.Vnonlin1.mgtr1} and \eqref{U.Vnonlin1.mlss1}; 
$n<1$ in cases \eqref{U.Vnonlin2.mgtr1} and \eqref{U.Vnonlin2.mlss1}; 
$1<n<2$ in case \eqref{U.Vnonlin3.mgtr1} 
and $m<n<m+1$ in case \eqref{U.Vnonlin3.mlss1}. 
\end{thm}

\begin{rem}
The quadratures in Theorem~\ref{thm:compactons.nonlin} 
overlap with the special case $m=n\neq1$, $k=0$
considered in Theorem~\ref{thm:compactons.lin}. 
This case can be excluded by imposing the condition that $B\neq0$ when $m=n$,
since $B\propto k$ from expression \eqref{B.A.kappa}. 
\end{rem}

There are numerous cases in which the quadratures can be evaluated 
explicitly in terms of elementary functions or elliptic functions.
The following specific types of symmetric solutions \eqref{cutoff.soln} will be considered:
\begin{subequations}\label{types}
\begin{align}
&\text{ cosine compactons }\quad
U_\c=\alpha \cos(\beta\xi)^q H(L-|\xi|),
\quad
L=\tfrac{1}{2\beta}\pi
\label{cos}\\
&\text{ Jacobi $\cn$ compactons }\quad
U_\c=\alpha\cn(\beta\xi,l)^q H(L-|\xi|),
\quad
L=\tfrac{1}{\beta}\K(l)
\label{cn}\\
&\text{ Jacobi $\sn$ compactons }\quad
U_\c=\alpha\sn(\beta(\xi+L),l)^q H(L-|\xi|),
\quad
L=\tfrac{1}{\beta}\K(l)
\label{sn}\\
&\text{ algebraic compactons }\quad
U_\c=\alpha \big(1-\beta\xi^2\big)^q H(L-|\xi|),
\quad
L=\tfrac{1}{\sqrt{\beta}}
\label{alg}
\end{align}
\end{subequations}
Here, $\alpha\neq 0$, $\beta>0$, $q>0$, $l\neq 0,1$ are constants,
while $\xi$ is the travelling wave variable \eqref{travellingwave};
$\K$ denotes the complete elliptic integral of the first kind. 

These types of solutions can be derived directly by substitution of
the given form for $U$ into the travelling wave ODE \eqref{U.ODE}, 
which yields a system of algebraic equations to be solved for the preceding constants 
$\alpha,\beta,q,l,m,n$,
with $a,b$ taken to be arbitrary.
This analysis has been carried out in Maple by 
using the command `rifsimp' to find all solutions.
In particular,
the Maple computation is complete for each type of compacton \eqref{types}.

The results are summarized in the next three subsections for the separate cases
$n=m\neq1$, $k\neq 0$;
$n\neq m\neq1$, $k=0$;
$n\neq m\neq1$, $k\neq 0$.
The quadrature family \eqref{U.Vnonlin1}, \eqref{U.Vnonlin2}, \eqref{U.Vnonlin3} 
to which each solution belongs will be stated (cf Table~\ref{table:properties}).

\begin{rem}
The case $n=1$ (linear dispersion)
belongs to the quadrature family \eqref{U.Vnonlin3},
which can be evaluated explicitly. 
It yields a cosine compacton \eqref{cos}
(cf expression \eqref{nis1.cos2}). 
\end{rem}

\subsection{Explicit solutions with $m=n$, $k\neq0$}\hfil

\emph{Cosine compactons}:

\begin{equation}\label{misn.cos1}
\begin{aligned}
& u = \big(\tfrac{2 n\k}{(n+1)a}\big)^{\frac{1}{n-1}}
\cos\Big( \tfrac{n-1}{2 n}\sqrt{\tfrac{a}{b}}\xi \Big)^{\frac{2}{n-1}} H(L-|\xi|),
\quad
L=\pi\tfrac{n}{n-1}\sqrt{\tfrac{b}{a}} ,
\\&
1<n< \tfrac{3}{2} ,
\quad
\sgn(\k)=\sgn(b)=\sgn(a).
\end{aligned}
\end{equation}
This solution generalizes the cosine compactons reported 
for the $K(2,2)$ and $K(3,3)$ equations in \Ref{RosHym}.
In particular, it reduces to dimension $N=1$ by putting $\s=0$,
whereby $\k = \nu$ is the wave speed.

\subsection{Explicit solutions with $m\neq n$, $k=0$}\hfil

\emph{Algebraic compactons}:

\begin{equation}\label{kis0.zsq3}
\begin{aligned}
& u =\alpha \Big(1 -\tfrac{a}{12 b\alpha^{\frac{n}{2}}} \xi^2\Big)^{\frac{2}{n}} H(L-|\xi|), 
\quad
L = \sqrt{\tfrac{12 b\alpha^{\frac{n}{2}}}{a}},
\\&
m = \tfrac{1}{2}n,
\quad
0 < n < \tfrac{1}{2},
\quad
\sgn(b)=\sgn(a) .
\end{aligned}
\end{equation}

\emph{Jacobi $\cn$ compactons}:

\begin{equation}\label{kis0.cn3}
\begin{aligned}
& u =\alpha \cn\Big(\sqrt{\tfrac{a\alpha^{2 n}}{b}} \xi, \tfrac{1}{\sqrt{2}}\Big)^{\frac{1}{n}} H(L-|\xi|), 
\quad
L= \sqrt{\tfrac{b}{a\alpha^{2 n}}} \K\big(\tfrac{1}{\sqrt{2}}\big) , 
\\&
m = 3 n,
\quad
0 < n < \tfrac{1}{4},
\quad
\sgn(b)=\sgn(a) .
\end{aligned}
\end{equation}

\begin{equation}\label{kis0.cn4}
\begin{aligned}
& u =\alpha \cn\Big(\sqrt{\tfrac{a\alpha^n}{3b}} \xi, \tfrac{1}{\sqrt{2}}\Big)^{\frac{2}{n}} H(L-|\xi|),
\quad
L=\sqrt{\tfrac{3b}{a\alpha^n}} \K\big(\tfrac{1}{\sqrt{2}}\big) , 
\\&
m = 2 n,
\quad
0 < n < \tfrac{1}{2} ,
\quad
\sgn(b)=\sgn(a) .
\end{aligned}
\end{equation}

\emph{Jacobi $\sn$ compactons}:

\begin{equation}\label{kis0.sn3}
\begin{aligned}
& u =\alpha \sn\Big(\sqrt{\tfrac{a\alpha^{2 n}}{2b}}(\xi+L), i\Big)^{\frac{1}{n}} H(L-|\xi|), 
\quad
L= \sqrt{\tfrac{2b}{a\alpha^{2 n}}} \K\big(i\big) , 
\\&
m = 3 n,
\quad
0 < n < \tfrac{1}{4},
\quad
\sgn(b)=\sgn(a) .
\end{aligned}
\end{equation}

\begin{equation}\label{kis0.sn4}
\begin{aligned}
& u =\alpha \sn\Big(\sqrt{\tfrac{a\alpha^n}{6b}}(\xi+L), i\Big)^{\frac{2}{n}} H(L-|\xi|),
\quad
L=\sqrt{\tfrac{6b}{a\alpha^n}} \K\big(i\big) , 
\\&
m = 2 n,
\quad
0 < n < \tfrac{1}{2} ,
\quad
\sgn(b)=\sgn(a) .
\end{aligned}
\end{equation}

\subsection{Explicit solutions with $m\neq n$, $k\neq0$}\hfil

\emph{Algebraic compactons}:

\begin{equation}\label{zsq1}
\begin{aligned}
& u = \big(\tfrac{k(3n+1)}{2 a (n+1)}\big)^{\frac{2}{n - 1}} \Big(1 - \tfrac{a^2(n+1)(n-1)^2}{2kbn(3n+1)^2}\xi^2\Big)^{\frac{2}{n-1}}  H(L-|\xi|),
\quad
L =\tfrac{3n+1}{(n-1)|a|} \sqrt{\tfrac{2n|kb|}{n+1}},
\\&
m = \tfrac{1}{2}(n+1),
\quad
1 < n< \tfrac{3}{2},
\quad
\sgn(k)=\sgn(a)=\sgn(b) . 
\end{aligned}
\end{equation}
This is a generalization of the polynomial compacton found for the $K(3,2)$ equation in \Ref{RosHym}. 

\begin{equation}\label{zsq2}
\begin{aligned}
&  u = \big(\tfrac{a (n+1)}{2 k}\big)^{\frac{1}{n-1}}
\Big( 1 -\tfrac{ k^2(n-1)^2}{-abn(n +1)^2} \xi^2\Big)^{\frac{1}{n - 1}} H(L-|\xi|),
\quad
L=\tfrac{(n+1)\sqrt{n|ab|}}{(n-1)|k|} , 
\\&
m = 2 -n,
\quad
1 < n < \tfrac{5}{4}, 
\quad
\sgn(k)=\sgn(a)=-\sgn(b) . 
\end{aligned}
\end{equation}

\emph{Cosine compactons}:
\begin{equation}\label{nis1.cos2}
\begin{aligned}
& u = \Big(\tfrac{2 a}{\k (m+1)}\Big)^{\frac{1}{1 - m}} \cos\Big(\sqrt{\tfrac{-k}{4b}} (1-m) \xi\Big)^{\frac{2}{1-m}a} H(L-|\xi|),
\quad
L = \pi \sqrt{\tfrac{|b|}{|k|}} (1 -m), 
\\&
n = 1,
\quad
\tfrac{1}{2} < m < 1,
\quad
\sgn(k)=\sgn(a) =-\sgn(b). 
\end{aligned}
\end{equation}
This is a generalization of the cosine compacton known \cite{PelSluKokPel}
for the sublinear gKdV equation. 

\emph{Jacobi cn compactons}:

\begin{equation}\label{cn1}
\begin{aligned}
& u = \Big(\tfrac{k(3 n - 1)}{a (n +1)}\Big)^{\frac{1}{2 n - 2}} 
\cn\big((1-n)\sqrt{\tfrac{-a}{nb}}\sqrt[4]{\tfrac{k}{a(3n-1)(n+1)}} \xi\big),
\\&
L=\tfrac{\sqrt{n|b|}}{1-n} \sqrt[4]{\tfrac{(3n-1)(n+1)}{ak}} K(\tfrac{1}{\sqrt{2}}),
\\
& m = 2 n - 1,
\quad
\tfrac{1}{2} < n < 1,
\quad
\sgn(k)=\sgn(a) =-\sgn(b). 
\end{aligned}
\end{equation}

\begin{equation}\label{cn2}
\begin{aligned}
& u = \Big(\tfrac{k(3 n - 1)}{a (n +1)}\Big)^{\frac{1}{2 n - 2}} 
\cn\Big((n-1)\sqrt{\tfrac{a}{bn}} \sqrt[4]{\tfrac{k}{a(3 n - 1)(n+1)}} \xi, \tfrac{1}{\sqrt{2}}\Big)^{\frac{2}{n - 1}} H(L-|\xi|),
\\& 
L=\tfrac{\sqrt{n|b|}}{n-1} \sqrt[4]{\tfrac{(3 n - 1)(n+1)}{ak}} \K\big(\tfrac{1}{\sqrt{2}}\big), 
\\&
m=2n-1 ,
\quad
1<n<\tfrac{3}{2},
\quad
\sgn(k)=\sgn(a) =\sgn(b) . 
\end{aligned}
\end{equation}

\emph{Jacobi $\sn$ compactons}:

\begin{equation}\label{sn1}
\begin{aligned}
& u = \big(\tfrac{k(3 n - 1)}{a (n +1)}\big)^{\frac{1}{2 n - 2}} 
\sn\Big((1-n)\sqrt{\tfrac{-a}{2bn}} \sqrt[4]{\tfrac{k}{a(3 n - 1)(n+1)}} (\xi+z0), i\Big)^{\frac{2}{1-n}} H(L-|\xi|),
\\& 
L=\tfrac{\sqrt{2n|b|}}{1-n} \sqrt[4]{\tfrac{(3 n - 1)(n+1)}{ak}} \K\big(i\big),
\\&
m=2n-1 ,
\quad
\tfrac{1}{2}<n<1,
\quad
\sgn(k)=\sgn(a) =-\sgn(b) . 
\end{aligned}
\end{equation}

\begin{equation}\label{sn2}
\begin{aligned}
& u = \big(\tfrac{k(3 n - 1)}{a (n +1)}\big)^{\frac{1}{2 n - 2}} 
 \sn\Big((n-1)\sqrt{\tfrac{a}{2bn}} \sqrt[4]{\tfrac{k}{a(3 n - 1)(n+1)}} (\xi+L), i\Big)^{\frac{2}{n - 1}} H(L-|\xi|),
\\& 
L=\tfrac{\sqrt{2n|b|}}{n-1}\sqrt[4]{\tfrac{(3 n - 1)(n+1)}{ak}} \K\big(i\big),
\\&
m=2n-1 ,
\quad
1<n<\tfrac{3}{2},
\quad
\sgn(k)=\sgn(a) =\sgn(b) . 
\end{aligned}
\end{equation}

\subsection{Antisymmetric compactons}\hfil

Each of the cosine, $\cn$ and $\sn$ compactons \eqref{misn.cos1}--\eqref{kis0.sn4} and \eqref{nis1.cos2}--\eqref{sn2}
has an anti-symmetric counterpart as follows
when the power $p$ is an odd (rational or integer) number. 

\emph{Sine compactons}:

\begin{equation}\label{antisymm.misn.sin1}
\begin{aligned}
& u = \big(\tfrac{2 n\k}{(n+1)a}\big)^{\frac{1}{n-1}}
\sin\Big( \tfrac{n-1}{2n}\sqrt{\tfrac{a}{b}}\xi \Big)^{\frac{2}{n-1}} H(L-|\xi|),
\quad
L=\pi\tfrac{2n}{n-1}\sqrt{\tfrac{2b}{a}} ,	
\\&
1<n<\tfrac{3}{2},
\quad
\sgn(\k)=\sgn(b)=\sgn(a).
\end{aligned}
\end{equation}

\begin{equation}\label{antisymm.nis1.sin2}
\begin{aligned}
& u = \big(\tfrac{(m+1)\k}{2a}\big)^{\frac{1}{m-1}}
\sin\Big((1-m)\sqrt{\tfrac{-k}{4b}}\xi \Big)^{\frac{2}{1-m}} H(L-|\xi|),
\\&
L=2\pi\tfrac{1}{1-m}\sqrt{\tfrac{|b|}{|k|}} ,
\\&
n=1,
\quad
\tfrac{1}{2}<m<1,
\quad
\sgn(\k)=\sgn(b)=\sgn(a).
\end{aligned}
\end{equation}

\emph{Jacobi $\sn$ compactons}:

\begin{equation}\label{antisymm.sn1}
\begin{aligned}
& u = \Big(\tfrac{k(3 n - 1)}{a (n +1)}\Big)^{\frac{1}{2 n - 2}} 
  \sn\Big((1-n)\sqrt{\tfrac{-a}{2bn}} \sqrt[4]{\tfrac{k}{a(3 n - 1)(n+1)}} \xi, i\Big)^{\frac{2}{1-n}} H(L-|\xi|),
\\& 
L=\tfrac{\sqrt{2n|b|}}{1-n} \sqrt[4]{\tfrac{(3n - 1)(n+1)}{ak}} \K\big(i\big),
\\&
m=2n-1 ,
\quad
\tfrac{1}{2}<n<1,
\quad
\sgn(k)=\sgn(a) = -\sgn(b) . 
\end{aligned}
\end{equation}

\begin{equation}\label{antisymm.sn2}
\begin{aligned}
& u = \big(\tfrac{k(3 n - 1)}{a (n +1)}\big)^{\frac{1}{2 n - 2}}
\sn\Big((n-1)\sqrt{\tfrac{a}{2nb}} \sqrt[4]{\tfrac{k}{a(3 n - 1)(n+1)}} \xi, i\Big)^{\frac{2}{n - 1}} H(L-|\xi|),
\\& 
L=\tfrac{\sqrt{2nb}}{n-1} \sqrt[4]{\frac{(3 n - 1)(n+1)}{ak}} \K\big(i\big), 
\\&
m=2n-1 ,
\quad
1<n<\tfrac{3}{2},
\quad
\sgn(k)=\sgn(a) =\sgn(b) . 
\end{aligned}
\end{equation}

\begin{equation}\label{antisymm.kis0.sn3}
\begin{aligned}
& u =\alpha \sn\Big(\sqrt{\tfrac{a\alpha^{2 n}}{2b}} \xi, i\Big)^{\frac{1}{n}} H(L-|\xi|), 
\quad
L= \sqrt{\tfrac{2b}{a\alpha^{2 n}}} \K\big(i\big) , 
\\&
m = 3 n,
\quad
0 < n < \tfrac{1}{4},
\quad
\sgn(b)=\sgn(a) .
\end{aligned}
\end{equation}

\begin{equation}\label{antisymm.kis0.sn4}
\begin{aligned}
& u =\alpha \sn\Big(\sqrt{\tfrac{a\alpha^n}{6b}} \xi, i\Big)^{\frac{2}{n}} H(L-|\xi|),
\quad
L=\sqrt{\tfrac{6b}{a\alpha^n}} \K\big(i\big) , 
\\&
m = 2 n,
\quad
0 < n < \tfrac{1}{2} ,
\quad
\sgn(b)=\sgn(a) .
\end{aligned}
\end{equation}

\emph{Jacobi $\cn$ compactons}:

\begin{equation}\label{antisymm.cn1}
\begin{aligned}
& u = \Big(\tfrac{k(3 n - 1)}{a (n +1)}\Big)^{\frac{1}{2 n - 2}} \cn\Big((1-n)\sqrt{\tfrac{-a}{bn}} \sqrt[4]{\tfrac{k}{a(3 n - 1)(n+1)}} (\xi+\tfrac{1}{2}L), \tfrac{1}{\sqrt{2}}\Big)^{\frac{2}{1-n}} H(L-|\xi|),
\\& 
L=\tfrac{2\sqrt{n|b|}}{1-n} \sqrt[4]{\tfrac{(3n - 1)(n+1)}{ak}} \K\big(\tfrac{1}{\sqrt{2}}\big),
\\&
m=2n-1 ,
\quad
\tfrac{1}{2}<n<1,
\quad
\sgn(k)=\sgn(a) = -\sgn(b) . 
\end{aligned}
\end{equation}

\begin{equation}\label{antisymm.cn2}
\begin{aligned}
& u = \big(\tfrac{k(3 n - 1)}{a (n +1)}\big)^{\frac{1}{2 n - 2}}
\cn\Big((n-1)\sqrt{\tfrac{a}{2nb}} \sqrt[4]{\tfrac{k}{a(3 n - 1)(n+1)}} (\xi+\tfrac{1}{2} L), \tfrac{1}{\sqrt{2}}\Big)^{\frac{2}{n - 1}} H(L-|\xi|),
\\& 
L=\tfrac{2\sqrt{n|b|}}{n-1} \sqrt[4]{\tfrac{(3 n - 1)(n+1)}{ak}} \K\big(\tfrac{1}{\sqrt{2}}\big), 
\\&
m=2n-1 ,
\quad
1<n<\tfrac{3}{2},
\quad
\sgn(k)=\sgn(a) =\sgn(b) . 
\end{aligned}
\end{equation}

\begin{equation}\label{antisymm.kis0.cn3}
\begin{aligned}
& u =\alpha \cn\Big(\sqrt{\tfrac{a\alpha^{2 n}}{b}} (\xi+\tfrac{1}{2}L), \tfrac{1}{\sqrt{2}}\Big)^{\frac{1}{n}} H(L-|\xi|), 
\quad
L= 2\sqrt{\tfrac{b}{a\alpha^{2 n}}} \K\big(\tfrac{1}{\sqrt{2}}\big) , 
\\&
m = 3 n,
\quad
0 < n < \tfrac{1}{4},
\quad
\sgn(b)=\sgn(a) .
\end{aligned}
\end{equation}

\begin{equation}\label{antisymm.kis0.cn4}
\begin{aligned}
& u =\alpha \cn\Big(\sqrt{\tfrac{a\alpha^n}{3b}} (\xi+\tfrac{1}{2}L), \tfrac{1}{\sqrt{2}}\Big)^{\frac{2}{n}} H(L-|\xi|),
\quad
L=2\sqrt{\tfrac{3b}{a\alpha^n}} \K\big(\tfrac{1}{\sqrt{2}}\big) , 
\\&
m = 2 n,
\quad
0 < n < \tfrac{1}{2} ,
\quad
\sgn(b)=\sgn(a) .
\end{aligned}
\end{equation}

\subsection{Kinematic features and profiles}

\begin{figure}[h!]
\centering
\includegraphics[trim=2cm 17cm 10cm 2cm,clip,width=0.45\textwidth]{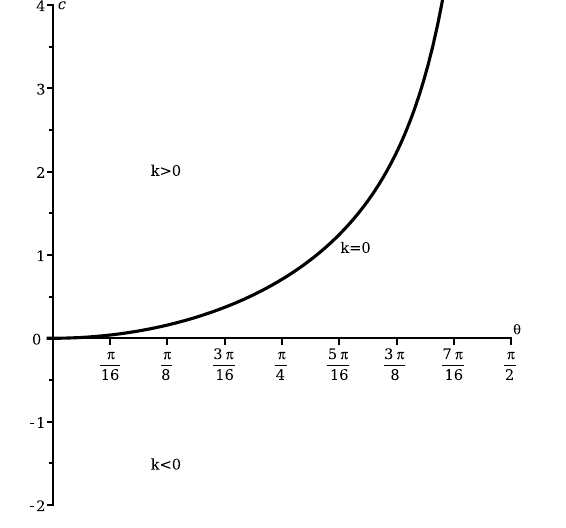}
\quad
\includegraphics[trim=2cm 17cm 10cm 2 cm,clip,width=0.45\textwidth]{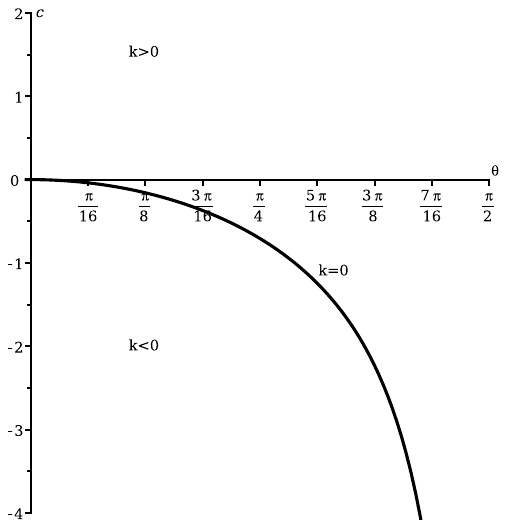}
\caption{
Allowed kinematic regions $(c,\theta)$:
$\sigma = 1$ (left); $\sigma = -1$ (right).
Only the portion $0\leq\theta<\tfrac{1}{2}\pi$ is shown
due to reflection symmetry. 
}\label{fig:kin-region}
\end{figure}

All of the line compactons \eqref{U.Vnonlin1}--\eqref{U.Vnonlin3} have
speed $c=\nu/\sqrt{1+\mu^2}$ and direction $\theta=\arctan(\mu)$
with respect to the positive $x$ axis.
In their profiles, $c$ and $\theta$ appear only in the combination 
\begin{equation}\label{k.c.theta}
\k=\nu -\s\mu^2 = \frac{c|\cos(\theta)|- \s\sin(\theta)^2}{\cos(\theta)^2} .
\end{equation}
The three possible cases $k=0$, $k>0$, $k<0$
determine different kinematic regions in $(c,\theta)$
as given by the relation \eqref{k.c.theta}.
These regions are shown in Fig.~\ref{fig:kin-region}.
Note that $\theta \to \theta \pm \pi$ is equivalent to $c \to -c$.
Also note that $\theta \neq \pm\tfrac{1}{2}\pi$ due to $|\mu|<\infty$,
i.e.\ purely transverse motion is disallowed. 
Hence, without loss of generality, $c$ will be taken to be positive or negative,
with $-\tfrac{1}{2}\pi <\theta <\tfrac{1}{2}\pi$.

Consequently, line compactons having different speeds and directions
but with the same $\k$ will share the same profile shape
and the same kinematic region. 
These shapes depend on 
the nonlinearity coefficients (convexities) $a$ and $b$,
the dispersion power $n>0$, and the convection power $m>0$, in addition to $k$.
A scaling of $t$, $x$, $y$ can be used to put $|a|=|b|=1$:
$t\to \sqrt{|b|/|a|^3}\,t$, 
$x\to \sqrt{|b|/|a|}\,x$,
$y\to \sqrt{|b|}/|a|\,y$.
In addition, a reflection in $x$ can then be used to put $a=1$ with $\s\to -\s$.
For ease of consideration of the physical behaviour,
only the scaling will be used hereafter.

Table~\ref{table:properties} summarizes some basic features of
the symmetric line compactons:
height and width, sign of $k$, nonlinearity coefficients (convexity),
and the dispersion power.
The antisymmetric counterparts of the cosine, cn and sn line compactons
have similar features. 

\begin{table}
\hbox{\hspace{-0.5in}
\begin{tabular}{c|c|c|c|l|l|l|l}
\hline
type & height $\propto$ & width $\propto$ & $\sgn\,k$ & convexity & dispersion & convection & quadrature 
\\
&&&&& power && family
\\
\hline
\hline
algebraic \eqref{kis0.zsq3}
&
$\alpha$
&
$\alpha^n$
&
$0$
&
$\sgn\,b=\sgn\,a$
&
$n<\tfrac{1}{2}$
&
sublinear
&
\eqref{U.Vnonlin2}, \eqref{U.Vnonlin2.mlss1}, \eqref{U.Vnonlin2.mlss1.Bis0}
\\
algebraic \eqref{zsq1}
&
$|k|^{\frac{2}{n-1}}$
&
$|k|^{\frac{1}{2}}$
&
$\sgn\, a$
&
$\sgn\,b=\sgn\,a$,
& 
$1<n<\tfrac{3}{2}$
&
superlinear
&
\eqref{U.Vnonlin3}, \eqref{U.Vnonlin3.mgtr1}, \eqref{U.Vnonlin3.mgtr1.Agtr0}
\\
algebraic \eqref{zsq2}
&
$1/|k|^{\frac{1}{n-1}}$
&
$1/|k|^{\frac{1}{2}}$
&
$\sgn\, a$
&
$\sgn\,b=-\sgn\,a$
& 
$1<n<\tfrac{5}{4}$
&
sublinear
&
\eqref{U.Vnonlin3}, \eqref{U.Vnonlin3.mlss1}, \eqref{U.Vnonlin3.mlss1.Blss0}
\\
\hline
cosine \eqref{misn.cos1}
&
$|k|^{\frac{1}{n-1}}$
&
--
&
$\sgn\, a$
&
$\sgn\,b=\sgn\,a$,
&
$1<n<\tfrac{3}{2}$
&
superlinear
&
\eqref{U.Vnonlin3}, \eqref{U.Vnonlin3.mgtr1}, \eqref{U.Vnonlin3.mgtr1.Agtr0}
\\
cosine \eqref{nis1.cos2} 
&
$1/|k|^{\frac{1}{1-m}}$
&
$1/|k|^{\frac{1}{2}}$
&
$\sgn\, a$
&
$\sgn\,b=\sgn\,a$
&
$n=1$
&
sublinear
&
\eqref{U.Vnonlin3}, \eqref{U.Vnonlin3.mlss1}, \eqref{U.Vnonlin3.mlss1.Blss0}
\\
\hline
cn \eqref{kis0.cn3} \& sn \eqref{kis0.sn3}
&
$\alpha$
&
$1/\alpha^{n}$
&
$0$
&
$\sgn\,b=\sgn\,a$,
&
$n<\tfrac{1}{4}$
&
sublinear
&
\eqref{U.Vnonlin1}, \eqref{U.Vnonlin1.mlss1}, \eqref{U.Vnonlin1.mlss1.Bis0.Agtr0}
\\
cn \eqref{kis0.cn4} \& sn \eqref{kis0.sn4}
&
$\alpha$
&
$1/\alpha^{\frac{n}{2}}$
&
$0$
&
$\sgn\,b=\sgn\,a$,
&
$n<\tfrac{1}{2}$
&
sublinear
&
\eqref{U.Vnonlin2}, \eqref{U.Vnonlin2.mlss1}, \eqref{U.Vnonlin2.mlss1.Bis0}
\\
cn \eqref{cn1} \& sn \eqref{sn1}
&
$1/|k|^{\frac{1}{2(1-n)}}$
&
$1/|k|^{\frac{1}{4}}$
&
$\sgn\, a$
&
$\sgn\,b=-\sgn\,a$,
&
$\tfrac{1}{2}<n<1$
&
sublinear
&
\eqref{U.Vnonlin3}, \eqref{U.Vnonlin3.mlss1}, \eqref{U.Vnonlin3.mlss1.Blss0}
\\
cn \eqref{cn2} \& sn \eqref{sn2} 
&
$|k|^{\frac{1}{2(n-1)}}$
&
$1/|k|^{\frac{1}{4}}$
&
$\sgn\, a$
&
$\sgn\,b=\sgn\,a$,
&
$1<n<\tfrac{3}{2}$
&
superlinear
&
\eqref{U.Vnonlin3}, \eqref{U.Vnonlin3.mgtr1}, \eqref{U.Vnonlin3.mgtr1.Agtr0}
\\
\hline
\end{tabular}
}
\caption{Basic features of the symmetric line compactons}
\label{table:properties}
\end{table}

\begin{figure}[h]
\centering
\includegraphics[trim=2cm 17cm 3cm 2cm,clip,width=0.75\textwidth]{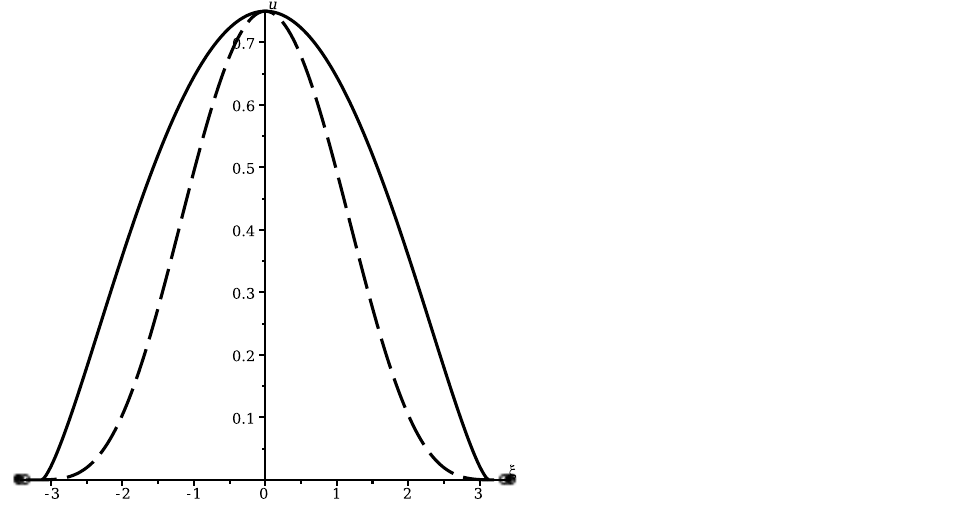}
\caption{Algebraic profile \eqref{kis0.zsq3}:
$n=\tfrac{1}{10}$ solid; $n=\tfrac{4}{9}$ dash.
Dots indicate $\pm L$.}
\label{fig:zsq3}
\end{figure}

\begin{figure}[h]
\centering
\includegraphics[trim=2cm 16cm 10cm 2cm,clip,width=0.48\textwidth]{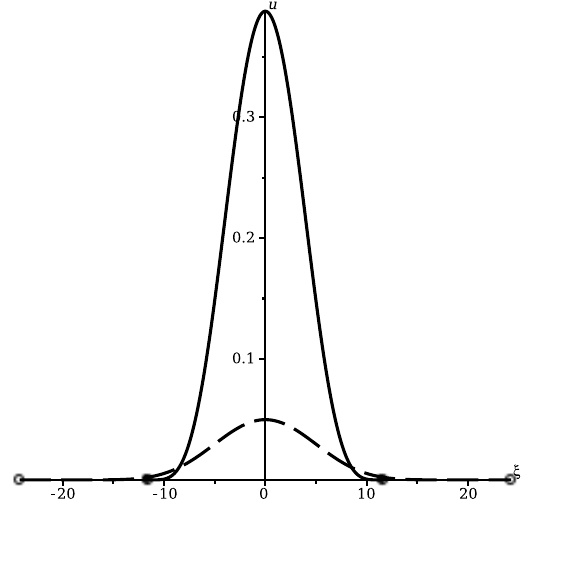}
\includegraphics[trim=2cm 16cm 10cm 2cm,clip,width=0.48\textwidth]{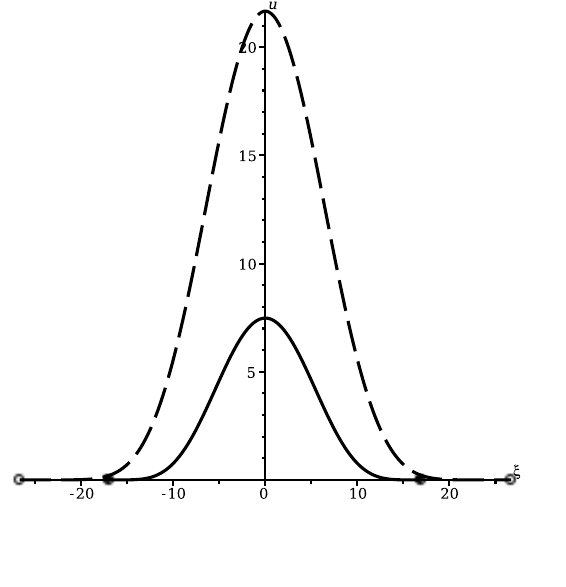}
\caption{
(left) Algebraic profile \eqref{zsq1}:
$n=\tfrac{10}{7}$ solid; $n=\tfrac{7}{6}$ dash.
(right) Algebraic profile \eqref{zsq2}:
$n=\tfrac{19}{16}$ solid; $n=\tfrac{10}{9}$ dash.
Dots indicate $\pm L$.
}\label{fig:zsq1-zsq2}
\end{figure}

Plots of the three algebraic line compactons are shown in
Figs.~\ref{fig:zsq3} and~\ref{fig:zsq1-zsq2}. 

\begin{figure}[h]
\centering
\includegraphics[trim=2cm 17cm 10cm 2cm,clip,width=0.45\textwidth]{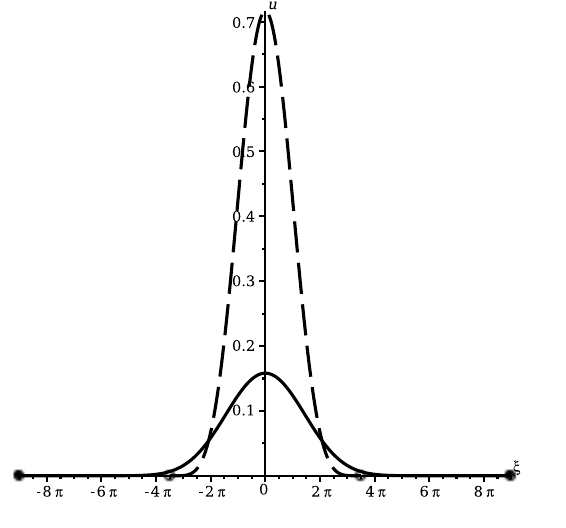}
\quad
\includegraphics[trim=2cm 17cm 10cm 2cm,clip,width=0.45\textwidth]{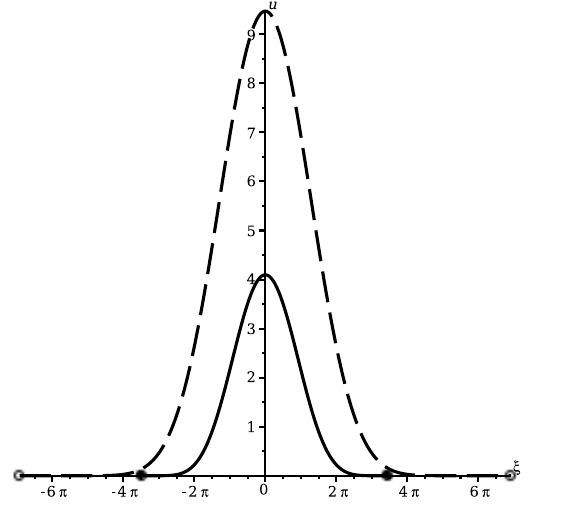}
\caption{
(left) Cosine profile \eqref{misn.cos1}:
$n=\tfrac{4}{3}$ solid; $n=\tfrac{6}{5}$ dash.
(right) Cosine profile \eqref{nis1.cos2}:
$m=\tfrac{2}{3}$ solid; $m=\tfrac{5}{6}$ dash.
Dots indicate $\pm L$.  
}\label{fig:cos1-cos2}
\end{figure}

\begin{figure}[h]
\centering
\includegraphics[trim=2cm 17cm 10cm 2cm,clip,width=0.45\textwidth]{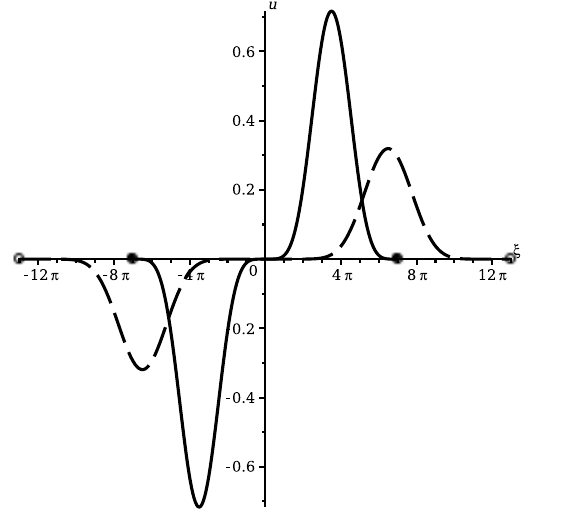}
\quad
\includegraphics[trim=2cm 17cm 10cm 2cm,clip,width=0.45\textwidth]{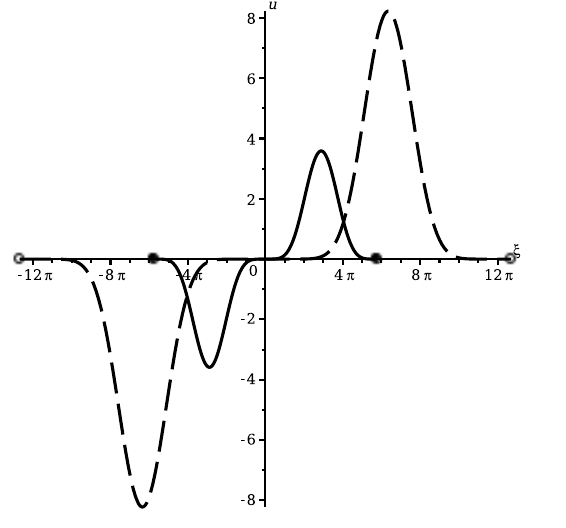}
\caption{
(left) Sine profile \eqref{antisymm.misn.sin1}:
$n=\tfrac{7}{5}$ solid; $n=\tfrac{13}{11}$ dash.
(right) Sine profile \eqref{antisymm.nis1.sin2}:
$m=\tfrac{3}{5}$ solid; $m=\tfrac{9}{11}$ dash.
Dots indicate $\pm L$.  
}\label{fig:sin1-sin2}
\end{figure}

Plots of the cosine and sine line compactons are shown in
Figs.~\ref{fig:cos1-cos2} and~\ref{fig:sin1-sin2}. 

\begin{figure}[h]
\centering
\includegraphics[trim=2cm 17cm 10cm 2cm,clip,width=0.48\textwidth]{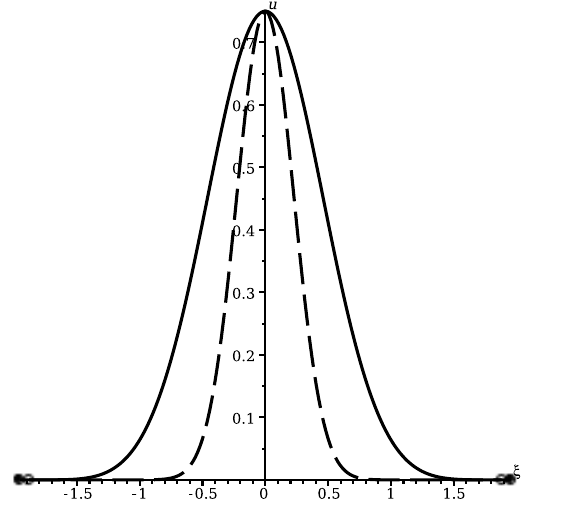}
\quad
\includegraphics[trim=2cm 17cm 10cm 2cm,clip,width=0.48\textwidth]{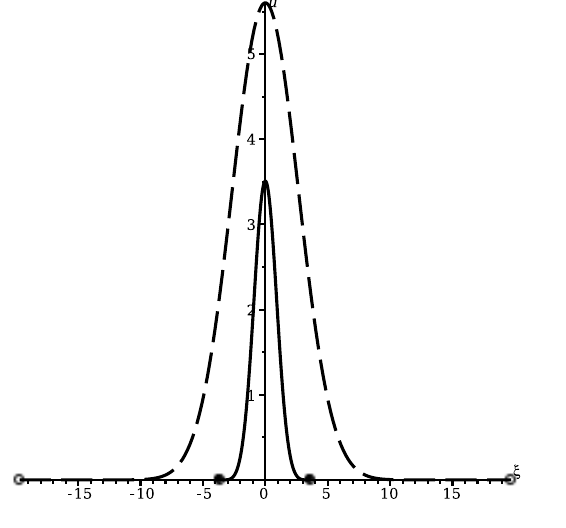}
\caption{
(left) cn profile \eqref{kis0.cn3}:
$n=\tfrac{3}{16}$ solid; $n=\tfrac{1}{20}$ dash.
(right) cn profile \eqref{cn1}:
$n=\tfrac{4}{7}$ solid; $n=\tfrac{7}{8}$ dash.
Dots indicate $\pm L$.
}\label{fig-cn3-cn1}
\end{figure}

Plots of two symmetric cn line compactons are shown in Fig.~\ref{fig-cn3-cn1}.
The other symmetric cn and sn line compacton have similar profiles. 

\begin{figure}[h]
\centering
\includegraphics[trim=2cm 17cm 10cm 2cm,clip,width=0.48\textwidth]{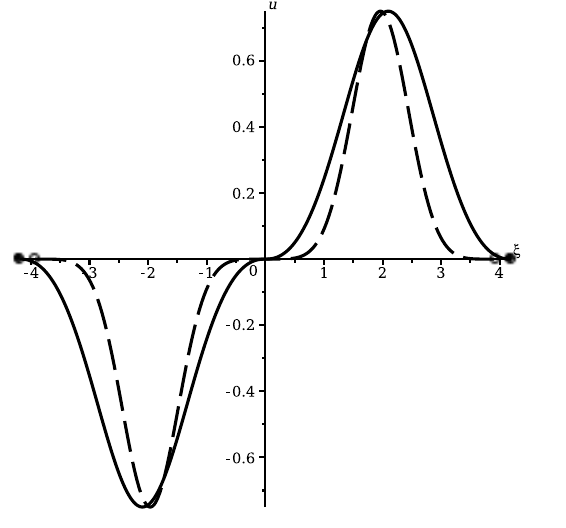}
\quad
\includegraphics[trim=2cm 17cm 10cm 2cm,clip,width=0.48\textwidth]{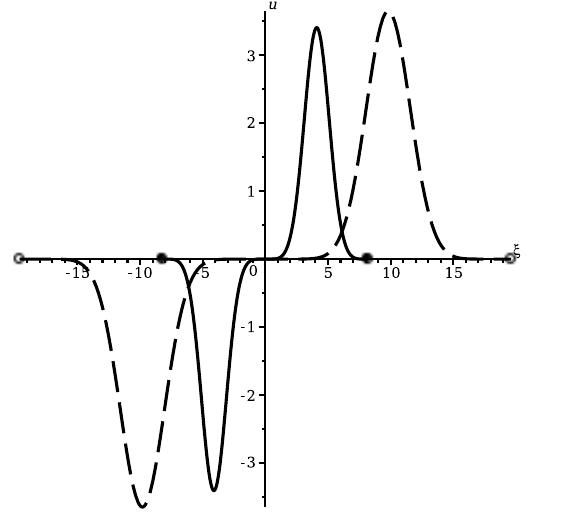}
\caption{
(left) sn profile \eqref{antisymm.kis0.sn3}:
$n=\tfrac{3}{7}$ solid; $n=\tfrac{1}{5}$ dash.
(right) sn profile \eqref{antisymm.sn1}:
$n=\tfrac{3}{5}$ solid; $n=\tfrac{7}{9}$ dash.
Dots indicate $\pm L$.
}\label{fig-sn3-sn1}
\end{figure}

Plots of two antisymmetric sn line compactons are shown in Fig.~\ref{fig-sn3-sn1}.
The other antisymmetric cn and sn line compacton have similar profiles. 

\section{Compactons in three (and higher) dimensions}\label{sec:Ndim}

The line compactons and solitary line waves obtained in the preceding sections
have a straightforward generalization to $N\geq 3$ dimensions,
which will now be presented.

To begin, observe that travelling wave solutions \eqref{3dim.travellingwave}
in $N=3$ dimensions
satisfy the fourth-order ODE
\begin{equation}\label{3dim.U.ODE}
( (\s|\mu|^2 -\nu)U+a U^m +b (U^n)'' )'' =0
\end{equation}
where $|\mu|=\sqrt{\mu_1{}^2+\mu_2{}^2}$.
This ODE has the same form \eqref{U.ODE} as in $N=2$ dimensions
with $\mu$ replaced by $|\mu|$.
As a consequence,
there is a direct correspondence between the solutions of the ODEs \eqref{U.ODE} and \eqref{3dim.U.ODE}.

Going to an arbitrary (higher) dimension $N>3$ leads to the same conclusion.
A plane travelling wave has the form $u=U(\xi)$ where, now,
$\xi = x +\mu_1 y_1 +\cdots + \mu_{N-1} y^{N-1} - \nu t$ is the travelling wave variable
in terms of Cartesian coordinates $(x,y^1,\ldots,y^{N-1})$ in $\Rnum^N$.
These travelling waves have constant amplitude $u$ in the hyperplane orthogonal to the direction of propagation.
Specifically, $u$ is invariant under the $N-1$ commuting translation symmetries
$\X_i = \mu_i \partial_x -\partial_{y^i}$, $i=1,\ldots,N-1$.
The ODE for $U(\xi)$ is unchanged from $N=3$ dimensions
apart from the expression $|\mu|^2=\sum_{i=1}^{N}\mu_i{}^2$ for $|\mu|$.

Thus, the dimension, $N\geq2$, does not play any significant role in either the nature or the behaviour of solutions.
The kinematical properties of plane travelling waves in $N\geq 3$ dimensions
are easiest to describe in a vector language
by writing the travelling wave variable as
$\xi = x+\vec{\mu}\cdot \vec{y}-\nu t$.
Then the vector $(1,\vec{\mu})$ is the direction of propagation in $\Rnum^N$,
while the speed is given by $c=\nu/\sqrt{1+\vec{\mu}\cdot\vec{\mu}}$.
The wave amplitude is constant in all directions orthogonal to the vector $(1,\vec{\mu})$.

\section{Conclusions}\label{sec:conclude}

Theorems~\ref{thm:compactons.lin} and~\ref{thm:compactons.nonlin}
provide an exhaustive classification of line compacton solutions
to the $K_N(m,n)$ equation in dimension $N=2$
and plane compacton solutions in dimensions $N\geq 3$.
In the case of linear dispersion, $n=1$,
a consequence of Theorem~\ref{thm:compactons.nonlin} is that
the only line compactons are given by the cosine and sine types \eqref{nis1.cos2} and \eqref{antisymm.nis1.sin2}, respectively.

All of the compacton solutions have counterparts satisfying the $K(m,n)$ equation,
which are given by taking $\s=0$ so that they reduce to dimension $N=1$,
where $k=\nu$ becomes the wave speed. 
In the case $k=0$, the resulting solutions are static. 

The KP equation, $K_2(2,1)$,
serves as a general model for different kinds of nonlinear waves that
exhibit weak nonlinearity and linear dispersion
and that also have a small transverse component. 
Physical applications include 
such as shallow water waves \cite{KadPet,AblSeg},
matter-wave pulses in Bose-Einstein condensates \cite{HuaMakVel},
ion-acoustic waves in plasmas \cite{InfRow}, ferromagnets \cite{KonDub},
and magnetic excitations in thin films \cite{VeeDan}. 
Other nonlinearities,
described by the $m$-power form of the $K_2(m,1)$ equation,
arise naturally in various nonlinear phenomena \cite{InfRow,KarBel,PelSteKiv}
and have been studied in analysis of the Cauchy problem for $m>1$ \cite{BouSau,WanAbl}.
The cosine compacton \eqref{nis1.cos2}, which has $n=1$ and $\tfrac{1}{2}<m<1$,
is applicable to models in which the convective nonlinearity is sublinear. 

The more general $K_2(m,n)$ equation with $n\neq1$
can be expected to arise in physical models with nonlinear dispersion
\cite{RosHymSta,RosZil2018}. 
The other compacton solutions we have obtained
will thereby have direct applicability in these models. 

Another important aspect of our results is that precise conditions are given
on the nonlinearity powers $m$ and $n$
under which a compacton expression is an actual solution of the $\KP_N(m,n)$ equation. 
A general statement of such conditions is often overlooked in the literature on compacton solutions to nonlinear equations. 
In fact,
often compacton ``solutions'' are derived by imposing only continuity at the cut off,
whereby the resulting expressions are not classical solutions,
namely they are purely formal piecewise solutions.
Such formal solutions are not physically meaningful, especially
in the context of evolution of initial data. 
See, e.g., \Ref{Yu,ZhaoQiaoTang} where this issue is overlooked for
nonlinearly dispersive KP-type equations.
Indeed, the $K_N(m,n)$ equation admits many formal piecewise solutions that are only continuous at the cut off,
which are not reported in the present paper because they fail to be classical solutions. 

Formal compacton ``solutions'' can be found in
\Ref{Waz05b,Sah,WeiTanChe,LiSon,YanBlu,Waz05a,Inc,MuDaiZhao,PinZhaMen}
for many nonlinearly dispersive versions of many nonlinear equations
originating in physics and applied mathematics.

An open problem of obvious relevance for understanding the role of compactons
in the nonlinear dynamics of physical systems
is to investigate whether compact initial data gives rise to trains of
compactons in the long-time limit as indicated in numerical experiments
(cf \Ref{ZilRos,RosZil2018}).

\section*{Acknowledgments}
SCA is supported by an NSERC Discovery grant.
MLG gratefully acknowledges support of Junta de Andalucia FQM-201 group.

\end{document}